\documentclass[12pt]{article}

\usepackage{amssymb,amsfonts,amsmath}
\usepackage{graphicx} 
\usepackage{color}
\usepackage{slashbox}
\usepackage[normalem]{ulem}
\usepackage{cancel}

\parskip=\medskipamount

\newcommand {\cC}{{\cal C}}

\newcommand {\cF}{{\cal F}}

\newcommand {\cH}{{\cal H}}

\newcommand {\cL}{{\cal L}}
\newcommand {\cM}{{\cal M}}
\newcommand {\cN}{{\cal N}}

\newcommand {\cR}{{\cal R}}

\def\a{\alpha}

\def\b{\beta}

\def\d{\delta}
\def\e{\epsilon}

\def\g{\gamma}
\def\G{\Gamma}

\def\k{\kappa}
\def\l{\lambda}

\def\q{\theta}

\def\s{\sigma}

\def\z{\zeta}

\def\F{\Phi}
\def\J{\Psi}
\def\L{\Lambda}
\def\O{\Omega}

\def\S{\Sigma}
\def\U{\Upsilon}

\newcommand{\ad}{{\dot{\alpha}}}                           
                            
\newcommand{\ve}{\varepsilon}

\newcommand{\pa}{\partial}                       
\newcommand{\hf}{\frac12}
\newcommand{\sect}[1]{\setcounter{equation}{0}\section{#1}}

\newcommand{\be}{\begin{equation}}
\newcommand{\ee}{\end{equation}}
\newcommand{\bea}{\begin{eqnarray}}
\newcommand{\eea}{\end{eqnarray}}
\newcommand{\non}{\nonumber}

\def\dt#1{{\buildrel {\hbox{\LARGE .}} \over {#1}}}

\def\double #1{#1{\hbox{\kern-2pt $#1$}}}

\newcommand{\oper}[1]{{\cal #1}}

\usepackage[normalem]{ulem}
\usepackage{cancel}
\begin{document}

\begin{titlepage}

\begin{flushright}
YGHP-14-02
\end{flushright}

\begin{center}
{\Large \bf Supersymmetry and Cotangent Bundle over}\\
{\vspace{3mm}}
{\Large \bf Non-compact Exceptional}\\
{\vspace{3mm}}
{\Large \bf Hermitian Symmetric Space}
\end{center}

\begin{center}
{\large  
$^{a}$Masato Arai\footnote{arai AT sci.kj.yamagata-u.ac.jp}, 
$^b$Kurando Baba\footnote{baba AT fukushima-nct.ac.jp}
} \\
\vspace{5mm}

{\it 
$^a$Faculty of Science, Yamagata University, Yamagata 990-8560, Japan\\
$^b$Department of General Education, \\
Fukushima National College of Technology, Fukushima 970-8034, Japan
}

\vspace{2mm}

\end{center}
\vspace{5mm}

\begin{abstract}
\baselineskip=14pt
\noindent
We construct $\cN=2$ supersymmetric nonlinear sigma models on the cotangent bundles over the non-compact exceptional Hermitian symmetric spaces $\cM=E_{6(-14)}/SO(10)\times U(1)$ and $E_{7(-25)}/E_6\times U(1)$. In order to construct them we use the projective superspace formalism which is an $\cN=2$ off-shell superfield formulation in four-dimensional space-time. This formalism allows us to obtain the explicit expression of $\cN=2$ supersymmetric nonlinear sigma models on the cotangent bundles over any Hermitian symmetric spaces in terms of the $\cN=1$ superfields, once the K\"ahler potentials of the base manifolds are obtained.
We derive the $\cN=1$ supersymmetric nonlinear sigma models on the K\"ahler manifolds $\cM$. Then we extend them into the $\cN=2$ supersymmetric models with the use of the result in arXiv:1211.1537 developed in the projective superspace formalism. The resultant models are the $\cN=2$ supersymmetric nonlinear sigma models on the cotangent bundles over the Hermitian symmetric spaces $\cM$. In this work we complete constructing the cotangent bundles over all the compact and non-compact Hermitian symmetric spaces.
\end{abstract}
\vspace{1cm}

\vfill
\end{titlepage}

\newpage
\setcounter{page}{1}
\renewcommand{\thefootnote}{\arabic{footnote}}
\setcounter{footnote}{0}

\tableofcontents{}
\vspace{1cm}
\bigskip\hrule

%%%%%%%%%%%%%%%%%%%%%%%%%%%%%%%%%%%%%%%%%%
%
% Section 1
%
%%%%%%%%%%%%%%%%%%%%%%%%%%%%%%%%%%%%%%%%%%
\sect{Introduction}
Supersymmetry (SUSY) has been studied in particle physics for a long time as a promising candidate of the theory beyond the Standard Model. On the other hand, it has been revealed that SUSY has an intimate relation to complex geometry in mathematics. Indeed, it is well known that target spaces of $\cN=1$ and $\cN=2$ SUSY nonlinear sigma models (NLSMs) must be K\"ahler \cite{Zumino} and hyperk\"ahler manifolds \cite{Alvarez-Gaume:1981hm}, respectively. It is important to construct these manifolds because they frequently appear in field theories with/without supersymmetry, supergravity and superstring theories. For instance,  moduli spaces of Bogomol'nyi-Prasad-Sommerfield monopoles are hyperk\"ahler manifolds \cite{AtHi}. The gravitational instanton which is the self-dual solution in the Euclidean Einstein's equation forms the hyperk\"ahler manifold \cite{EgHa}. The low energy effective actions of the $\cN=2$ SUSY gauge theories in the Higgs branch are the NLSMs on the hyperk\"ahler manifolds \cite{ArPl, AnPi}. The NLSMs on hyperk\"ahler manifolds have been intensively studied since they involve a lot of solitons such as domain walls, lumps, etc (for review, for instance, see \cite{EtIsNiOhSa}). 
Although hyperk\"ahler manifolds appear in various fields in particle physics, it is difficult to construct them because of their complex symmetry.

Recently there have been developments to construct $\cN=2$ SUSY NLSMs in the projective superspace formalism \cite{Karlhede:1984vr, Hitchin:1986ea, LR1, LR2, G-RLRWvU}, which is an $\cN=2$ off-shell superfield formulation in four-dimensional space-time\footnote{There is the harmonic superfield formalism \cite{GaIvKaOgSo} as another $\cN=2$ off-shell superfield formulation.}. 
In this formalism, $\cN=2$ SUSY NLSMs on cotangent bundles over K\"ahler  manifolds have been constructed \cite{GK1, GK2, ArNi, AKL, AKL2, KuNo, AB}.  A key observation in the developments is that once a certain $\cN=1$ SUSY NLSM is obtained, this model can be extended into the $\cN=2$ SUSY NLSM with use of the projective superspace formalism. If we have the $\cN=1$ SUSY NLSM on the K\"ahler manifold, we can obtain $\cN=2$ SUSY NLSM on the cotangent bundle over the K\"ahler manifold. The target space of the $\cN=2$ SUSY NLSM is shown to be an open domain of the zero section of the cotangent bundle \cite{GK1, GK2}. Namely it is hyperk\"ahler. There is also a similar development in mathematics: It is proved that for a K\"ahler manifold $\cM$ a K\"ahler structure on an neighborhood of the zero section of the cotangent bundle over $\cM$ exists \cite{Ka1,Ka2,Fe}. The proofs in \cite{GK1, GK2}  and \cite{Ka1,Ka2,Fe} were performed independently. Based on the observation in \cite{GK1, GK2}, the $\cN=2$ SUSY NLSMs on the cotangent bundles over the irreducible Hermitian symmetric spaces (HSSs) except the non-compact exceptional types of them have been constructed \cite{GK1, GK2, ArNi, AKL, AKL2, KuNo, AB}. The irreducible HSSs classified by Cartan \cite{cartan} consist of compact type and non-compact type. They are summarized in Table \ref{HSS}. 

Let us see more in detail about the recent developments of construction of $\cN=2$ SUSY NLSM.
The projective superspace consists of the standard ${\cal N}=2$ superspace and the so-called 
projective coordinate $\zeta$ parameterizing $SU(2)_R/U(1)$, where the $SU(2)_R$ is an internal 
symmetry of the ${\cal N}=2$ SUSY algebra. Superfields are defined on a subspace of the projective 
superspace. While there are several superfields on the subspace, a relevant one for
constructing SUSY NLSMs is called the polar multiplet which represents hypermultiplet. 
Starting with an $\cN=1$ SUSY NLSM on a K\"ahler manifold $\cM$ in terms of chiral
superfields, the $\cN=2$ SUSY extension is obtained by replacing the chiral superfields 
with the polar multiplets and integrating over the subspace of the projective 
superspace \cite{GK1, GK2}. 
The model obtained is written by fields representing the base manifold coordinates (chiral 
superfields) and tangent vectors (complex linear superfields) along with an infinite set
of unconstrained auxiliary superfields. After eliminating the auxiliary fields 
the model represents the $\cN=2$ SUSY model on the tangent bundle over ${\cal M}$, which
we write $S_{\rm tb}$.
Exchanging the tangent vectors in $S_{\rm tb}$ to the cotangent vectors with the aid of the generalized
Legendre transformation \cite{LR1}, we obtain the $\cN=2$ SUSY NLSM on the cotangent bundle 
$S_{\rm ctb}$ over $\cM$.

One of the main problems to obtain $\cN=2$ SUSY NLSMs in the above method is to solve the equations 
of motion for the auxiliary fields. While various ways to solve them have been proposed in constructing 
the $\cN=2$ SUSY NLSMs on cotangent bundle over the HSSs \cite{GK1, GK2, ArNi, AKL, AKL2, KuNo, AB}, 
the useful one is to use the properties of the HSS and SUSY \cite{AKL2}. This is applicable 
to all the HSSs. 
The way in \cite{AKL2} allows us to derive the general formula for the tangent bundle over the HSSs.
Another problem is to perform the generalized Legendre transformation.
It is technically difficult to perform it for all the HSSs to obtain the cotangent bundle from the tangent bundle.
It is easy to implement for all the compact and non-compact classical types of HSSs \cite{GK1, GK2, AKL} 
and the compact exceptional HSS $E_6/SO(10)\times U(1)$ \cite{AKL2}. However, it is difficult for the 
compact exceptional HSS $E_7/E_6\times U(1)$.
Accordingly another method has been developed \cite{AB} based on the result in \cite{KuNo}.
In \cite{KuNo}, the general form of the cotangent bundle over any HSS is derived. 
Afterwards, a different form is obtained in \cite{AB} and it has been applied to derivation of the $\cN=2$ 
SUSY NLSM on the cotangent bundle over the compact exceptional HSS $E_7/E_6\times U(1)$.

\begin{table}
\begin{center}
\begin{tabular}{c|cccc|cc}
 & \multicolumn{4}{c|}{classical type} & \multicolumn{2}{c}{exceptional type} \\ \hline
compact type & $U(n+m) \over U(n) \times U(m)$ & $SO(2n)\over U(n)$ & {$Sp(n)\over U(n)$} & $SO(n+2)\over SO(n)\times U(1)$ & $E_6 \over SO(10)\times U(1)$ & $E_7\over E_6\times U(1)$ \\ \hline
non-compact type & $U(n,m) \over U(n) \times U(m)$ & $SO^*(2n)\over U(n)$ & $Sp(n,{\bf R})\over U(n)$ &$SO_0(n,2)\over SO(n)\times U(1)$ & $E_{6(-14)} \over SO(10)\times U(1)$ & $E_{7(-25)}\over E_6\times U(1)$ 
\end{tabular}\label{HSS}
\end{center}
\caption{Irreducible Hermitian symmetric spaces}
\end{table}

In this paper we construct the $\cN=2$ SUSY NLSMs on cotangent bundles over the non-compact exceptional HSSs, 
${\cal M}=E_{6(-14)}/SO(10)\times U(1)$ and $E_{7(-25)}/E_6\times U(1)$ \footnote{$E_{6(-14)}$ denotes the real form of $E^{\bf C}_{6}$ with character $-14$ where ${\rm dim}(E_{6(-14)}/SO(10)\times U(1))-{\rm dim}(SO(10)\times U(1))$ gives $-14$. Similarly, $E_{7(-25)}$ is also defined.}
by using the results for the HSSs elaborated in \cite{KuNo} and \cite{AB}.
To this end, we need to derive the K\"ahler manifolds ${\cal M}$, which have not been known yet.
With the explicit forms of the K\"ahler potentials of $\cM$,  we apply the result in \cite{AB}, where 
a more useful formula for the action of the cotangent bundle over any HSSs is derived, and obtain the 
$\cN=2$ SUSY NLSMs on the cotangent bundles over ${\cal M}$.
In this work we complete constructing the $\cN=2$ SUSY NLSMs on the cotangent bundles over {\it all} the compact and non-compact HSSs.

The paper is organized as follows. In Section 2, we give a brief review of the ${\cal N}=2$ NLSMs in the projective superspace formalism. We also explain how to derive the explicit forms of the tangent bundle and the cotangent bundle actions for the HSSs.
In Section 3, we explain the $E_{6(-14)}$ algebra and then derive $\cN=1$ SUSY NLSM on $E_{6(-14)}/SO(10)\times U(1)$. 
We apply the result explained in Section 2 to derive the $\cN=2$ SUSY NLSM model of the cotangent bundle over $E_{6(-14)}/SO(10)\times U(1)$ and obtain it. 
In Section 4, we give the $E_{7(-25)}$ algebra and construct the $\cN=1$ SUSY model on $E_{7(-25)}/E_6\times U(1)$. Then we construct the $\cN=2$ SUSY NLSM model on the cotangent bundle over $E_{7(-25)}/E_6\times U(1)$ by applying the result in Section 2. Section 5 is devoted to conclusion. In Appendix \ref{appendix}, we briefly summarize the Clifford algebra used in Section 3. In Appendix \ref{calc}, detailed calculation to obtain (\ref{trans4}) from (\ref{trans3}) is summarized. In Appendix \ref{GLT}, another derivation of the $\cN=2$ SUSY NLSM on the cotangent bundle over $E_{6(-14)}/SO(10)\times U(1)$ is explained. There we construct the tangent bundle action by using the result in Section 2. Then we utilize the Legendre transformation and obtain the cotangent bundle over $E_{6(-14)}/SO(10)\times U(1)$. In Appendix D, derivation of the tangent bundle over $E_{7(-25)}/E_6\times U(1)$ using the result in Section 2 is explained.

%%%%%%%%%%%%%%%%%%%%%%%%%%%%%%%%%%%%%%%%%%
%
% Section 2
%
%%%%%%%%%%%%%%%%%%%%%%%%%%%%%%%%%%%%%%%%%%
\sect{$\cN=2$ sigma models and the projective superspace}
\subsection{General case}
Projective superspace is described as $(x_\mu,\theta_{\alpha i}, {\bar{\theta}}_{\dot{\alpha}}^i, \zeta)$, where $\mu=0,1,2,3$ is a space-time index, $\alpha, \dot{\alpha}=1,2$ are spinor indices, 
$i=1,2$ is an $SU(2)_R$ index and $\z$ is the projective coordinate.
Superfields are functions on its subspace, which are defined by the so-called projective condition \cite{Karlhede:1984vr} similar to the chiral condition in the four-dimensional ${\cal N}=1$ SUSY field theory. This condition makes a number of the Grassmann coordinates be half and integration measure for SUSY invariant action reduces to one on the full ${\cal N}=1$ superspace 
$z_M=(x_\mu,\theta_\alpha,\bar{\theta}_{\dot{\alpha}})$
with the projective coordinate $\z$. 
A certain class of four-dimensional ${\cal N}=2$ NLSM is described in terms of ${\cal N}=1$ language as \cite{Ku, GK1, GK2}
\begin{eqnarray}
 S[\U, \breve{\U}]  =  
 \frac{1}{2\pi i} \, \oint \frac{{\rm d}\z}{\z} \,  
 \int {\rm d}^8 z \, 
 K \big( \U^I (z, \z), \breve{\U}^{\bar{J}} (z, \z)  \big),
 \label{PSS-e1}
\end{eqnarray}
where $I, J$ are indices of fields\footnote{More general type of action has the form $K(\U,\breve{\U},\z)$ \cite{LR1, LR2}.}. 
The contour encircles the origin of the $\z$-plane in anti-clockwise direction.
The action is written by the function of the superfields representing
the polar multiplets $\U$ and $\breve{\U}$, which are called an arctic superfield and an antarctic superfield respectively.
They are expanded with respect to $\z$ as
\begin{eqnarray}
\U(z, \z) = \sum_{n=0}^{\infty}  \, \U_n \z^n = 
\F + \S \,\z+ {\cal A} ~,\qquad
\breve{\U} (\z) = \sum_{n=0}^{\infty}  \, {\bar
\U}_n (-\z)^{-n}\,, \label{PSS-e2}
\end{eqnarray}
where $\Upsilon_0\equiv \Phi$ is a chiral superfield ($\bar{D}_{\dot{\alpha}}\Phi=0$) and $\U_1=\Sigma$ is a complex linear superfield ($\bar{D}^2\Sigma=0$),
where $\bar{D}_{\dot{\a}}=-{\partial \over \partial \bar{\theta}^{\dot\a}}-i\theta^\alpha\sigma^\mu_{\a\dot{\alpha}}{\partial \over \partial x^\mu}$ is the covariant derivative in the $\cN=1$ superspace.
An infinite set of unconstrained auxiliary fields is expressed as ${\cal A}$ which contains terms with an order higher than $\zeta$.
The antarctic superfield $\breve{\U}$ is a conjugate of $\U$, which is the combination of the ordinary complex conjugate and the antipodal map $\z\mapsto -1/\z$ on the Riemann sphere.

The action (\ref{PSS-e1}) is an ${\cal N}=2$ extension of the general ${\cal N}=1$ SUSY NLSM \cite{Zumino}
\begin{eqnarray}
S[\F, \bar \F] =  \int {\rm d}^8 z \, K(\Phi^{I}, {\bar \Phi}{}^{\bar{J}})\,,
\label{PSS-e3}
\end{eqnarray}
where $K(\Phi^{I}, {\bar \Phi}{}^{\bar{J}})$ is the K\"ahler potential of a K\"{a}hler manifold.
Indeed, expanding (\ref{PSS-e1}) with respect to $\z$, we can see the K\"ahler potential in 
(\ref{PSS-e3}) is included \cite{GK1}:
\begin{eqnarray}
S={1 \over 2\pi i}\oint {d\z \over \z}\int d^8z e^{{\cal A}\partial+\breve{\cal A}\bar{\partial}}\exp\left(\z \S\partial-{1\over \z}\bar{\S}\bar{\partial}\right)K(\F^I,\bar{\F}^{\bar{J}})\,,
\end{eqnarray}
where $\partial\equiv \partial/\partial\Phi$ and $\bar{\partial}\equiv \partial/\partial \bar{\Phi}$, 
and the exponential factors in the integrand are the $\cN=2$ completion.
The action (\ref{PSS-e3}) is invariant under the K\"ahler transformation
\begin{eqnarray}
 K(\Phi^I, \bar{\Phi}^{\bar{J}})\rightarrow K(\Phi^I, \bar{\Phi}^{\bar{J}})+\Lambda(\Phi^I)+\bar{\Lambda}(\bar{\Phi}^{\bar{J}}), \label{KT}
\end{eqnarray}
and reparametrization of the manifold
\begin{eqnarray}
 \Phi^I \rightarrow f^I(\Phi^J). \label{RP}
\end{eqnarray}
In the ${\cal N}=2$ SUSY NLSM (\ref{PSS-e1}), 
(\ref{KT}) and (\ref{RP}) correspond to
\begin{eqnarray}
 K(\U^I, \breve{\U}^{\bar{J}})\rightarrow K(\U^I, \breve{\U}^{\bar{J}})+\Lambda(\U^I)+\breve{\Lambda}(\breve{\U}^{\bar{J}}),
\end{eqnarray}
and 
\begin{eqnarray}
 \U^I \rightarrow f^I(\U^J), \label{PSS-e4}
\end{eqnarray}
respectively.
From the transformation (\ref{PSS-e4}), we find the transformation law for $\Sigma^I$ as
\begin{eqnarray}
 \Sigma^I={d \U^I \over d\z}{\Bigg |}_{\z=0}\rightarrow {d f^I \over d\z}{\Bigg |}_{\z=0}={d \U^J \over d\z}{d f^I \over d\U^J}{\Bigg |}_{\z=0}=\Sigma^J{d f^I \over d\U^J}{\Bigg |}_{\z=0}.
\end{eqnarray}
It is seen that $\Sigma^I$ transforms as a tangent vector. 

In order to represent the action (\ref{PSS-e1}) in terms of physical fields $(\Phi^I, \Sigma^J)$ only, we need to eliminate the auxiliary fields by using their equations of motion
\begin{eqnarray}
\oint \frac{{\rm d} \z}{\z} \,\z^n \, \frac{\pa K(\U, \breve{\U} 
) }{\pa \U^I} ~ = ~ \oint \frac{{\rm d} \z}{\z} \,\z^{-n} \, \frac{\pa 
K(\U, \breve{\U})} {\pa \breve{\U}^{\bar I}}=0, \qquad n \geq 2 \,. \label{EoMA}
\end{eqnarray}
Let $\U_*(\z)\equiv\U_*(\z;\F,\bar{\F},\S,\bar{\S})$ be a unique solution of 
the equation (\ref{EoMA}) with the initial conditions
\begin{eqnarray}
 \U_*(0)=\Phi,\qquad {d \U(\z) \over d\z}{\Bigg |}_{\z=0}=\Sigma.
\end{eqnarray}
For a general K\"ahler manifold, it is possible to eliminate $\U_n(n\ge 2)$ and 
their conjugates by solving (\ref{EoMA}) perturbatively \cite{KuLi}. 
After eliminating all the auxiliary fields, the following form of the action is obtained
\begin{eqnarray}
S_{{\rm tb}}[\F,  \S]  
&=& \int {\rm d}^8 z \, \Big\{\,
K \big( \F, \bar{\F} \big)+ \cL\big(\F, \bar \F, \S , \bar \S \big) \Big\}\,, \label{act-tab}
\end{eqnarray}
where $\cL(\F, \bar \F, \S , \bar \S)$ is the part describing the tangent space:
\begin{eqnarray}
\cL\big(\F, \bar \F, \S , \bar \S \big)
&=& \sum_{n=1}^{\infty} \cL^{(n)}
\big(\F, \bar \F, \S , \bar \S \big) \non \\
&=&\sum_{n=1}^{\infty} \cL_{I_1 \cdots I_n {\bar J}_1 \cdots {\bar 
J}_n }  \big( \F, \bar{\F} \big) \S^{I_1} \dots \S^{I_n} 
{\bar \S}^{ {\bar J}_1 } \dots {\bar \S}^{ {\bar J}_n }. \label{tan}
\end{eqnarray}
Here $\cL_{I\bar{J}}=-g_{I\bar{J}}(\F,\bar{\F})$ while the tensors $\cL_{I_1 \cdots I_n 
{\bar J}_1 \cdots {\bar J}_n }(n\ge 2)$ are functions of the metric $g_{I\bar{J}}(\F,\bar{\F})$, the Riemann tensor 
$R_{I\bar{J}K\bar{L}}(\F, \bar{\F})$ and its covariant derivative. 
Each term of the action contains equal powers of $\S$ and $\bar{\S}$ because the action (\ref{PSS-e1}) possesses the $U(1)$ invariance \cite{GK1}
\begin{eqnarray}
 \U(\z)\rightarrow \U(e^{i\alpha}\z) \quad \Leftrightarrow \quad \U_n(z)\rightarrow e^{in\alpha}\U_n(z).
\end{eqnarray}
The action (\ref{act-tab}) is written by the base manifold coordinate $\F$ and the tangent 
vector $\S$. 
Therefore this action represents the $\cN=2$ SUSY model on
the tangent bundle over the K\"ahler manifold.

The rest of the work is to derive the K\"ahler potential of the cotangent bundle over
the K\"ahler manifold.
It is carried out by changing the tangent vectors
$\S$'s in (\ref{act-tab}) into chiral one-forms, cotangent vectors $\Psi$'s. 
It can be performed by the generalized Legendre transformation \cite{LR1} as follows. 
\begin{eqnarray}
&& S_{\rm tb}=\int d^8z \left(K(\Phi,\bar{\Phi})+\cL(\F,\bar{\F},\S,\bar{\S})\right) \nonumber \\
&& \leadsto S =  \int {\rm d}^8 z \, \Big\{\,
K \big( \F, \bar{\F} \big)+  \cL
\big(\F, \bar \F, U , \bar U \big)
+\J_I U^I + {\bar \J}_{\bar I} {\bar U}^{\bar I} 
\Big\}\,,
\label{f-o}
\end{eqnarray}
where $U$ is a complex unconstrained superfield and $\J$ is a chiral superfield.
This action goes back to the tangent bundle action (\ref{act-tab}) after eliminating the chiral superfields $\Psi$ and $\bar{\Psi}$ by their equations of motion. 
Indeed, the equations of motion for $\Psi$ and $\bar{\Psi}$ give the constraint $D^2 U=\bar{D}^2U=0$ which is exactly the condition for a complex linear superfield. On the other hand, eliminating $U$ and $\bar{U}$ with the aid of their equations of motion, the action is written only in terms of $\Phi, \Psi$ and their conjugates:
\begin{eqnarray}
S_{{\rm ctb}}[\F,  \J]  
&=& \int {\rm d}^8 z \, \Big\{\,
K \big( \F, \bar{\F} \big)+    
\cH \big(\F, \bar \F, \J , \bar \J \big)\Big\}\,,
\label{act-ctb}
\end{eqnarray}
where 
\begin{eqnarray}
\cH \big(\F, \bar \F, \J , \bar \J \big)&=& \sum_{n=1}^\infty \cH^{(n)}(\Phi,\bar{\Phi},\Psi, \bar{\Psi}) \nonumber \\
&=&\sum_{n=1}^{\infty} \cH^{I_1 \cdots I_n {\bar J}_1 \cdots {\bar 
J}_n }  \big( \F, \bar{\F} \big) \J_{I_1} \dots \J_{I_n} 
{\bar \J}_{ {\bar J}_1 } \dots {\bar \J}_{ {\bar J}_n }\,, \label{h}
\end{eqnarray}
with $\cH^{I {\bar J}} \big( \F, \bar{\F} \big)=g^{I {\bar J}} \big( \F, \bar{\F} \big)$. 
Here $g^{I {\bar J}}$ is the inverse metric of $g_{I {\bar J}}$.
The variables $(\F^I, \Psi_J)$ parameterize the cotangent bundle over the K\"ahler manifold and therefore the action gives the K\"ahler potential of the cotangent bundle over the K\"ahler manifold.

%%%%%%%%%%%%%%%%%%%%%%%%%%%%%%%%%%%%%%%%%%%%%%%%%%
%
% Hermitian symmetric space
%
%%%%%%%%%%%%%%%%%%%%%%%%%%%%%%%%%%%%%%%%%%%%%%%%%%
\subsection{HSS case}
The explicit forms of $\cL$ \cite{AKL2} and $\cH$ \cite{KuNo,AB} are obtained for the case that the base manifold is a HSS.
In this subsection we explain their derivations.

First we explain the derivation of the tangent space part $\cL$.
Since the tangent space part (\ref{tan}) hides the second SUSY invariance after 
eliminating the auxiliary fields perturbatively, we first derive the condition so that 
the $\cL$ is invariant under the second SUSY transformation.
Here we take into account
that the base manifold is the HSS. 
Second we solve the condition and obtain the explicit formula for $\cL$.

By construction in the projective superspace formalism, the action is invariant under the $\cN=2$ SUSY transformation \cite{LR1, LR2}
\begin{eqnarray}
\d \U^I (\z)= i \left(\ve^\a_i Q^i_\a +
{\bar \ve}^i_\ad {\bar Q}^\ad_i \right)  \U^I(\z),
\label{SUSY1}
\end{eqnarray}
where $\ve^\a_i, {\bar \ve}^i_{\dot{\a}}$ are transformation parameters and 
$Q^i_\a, {\bar Q}^\ad_i$ are supercharges. 
The action (\ref{act-tab}) is written in terms of $\cN=1$ superfield and therefore 
$\cN=1$ SUSY is only manifest.
In the following we investigate the invariance of the action under the second SUSY
transformation (\ref{SUSY1}) for $i=2$.
It is shown that the second SUSY acts on $\F$ and $\S$ as \cite{G-RLRWvU}
\begin{eqnarray}
\d \F^I = {\bar \ve}_{\dt{\a}} {\bar D}^{\dt{\a}} \S^I\,, \qquad 
\d \S^I = -\ve^\a D_\a \F^I +   {\bar \ve}_{\dt{\a}} {\bar D}^{\dt{\a}} \U_2^I\,. \label{SUSY0}
\end{eqnarray}
The condition for the Riemann tensor $R_{I_1 {\bar  J}_1 I_2 {\bar J}_2}$ of the HSS
\begin{eqnarray}
\nabla_L  R_{I_1 {\bar  J}_1 I_2 {\bar J}_2}
= {\bar \nabla}_{\bar L} R_{I_1 {\bar  J}_1 I_2 {\bar J}_2} =0\,,
\label{covar-const}
\end{eqnarray}
allows us to rewrite (\ref{SUSY0}) as
\begin{eqnarray}
\d \F^I = {\bar \ve}_{\dt{\a}} {\bar D}^{\dt{\a}} \S^I\,, \qquad 
\d \S^I = -\ve^\a D_\a \F^I -\hf   {\bar \ve}_{\dt{\a}} {\bar D}^{\dt{\a}} 
\Big\{ \G^I_{~JK} \big( \F, \bar{\F} \big) \, \S^J\S^K \Big\},
\label{SUSY2}
\end{eqnarray}
where $\nabla_L$ and $\bar{\nabla}_{\bar{L}}$ are covariant derivatives with respect to $\F$ and $\bar{\F}$, and $\Gamma^I_{JK}$ is the Christoffel symbol.
We impose the tangent bundle action (\ref{act-tab}) to be invariant under (\ref{SUSY2}), and then find \cite{AKL2}
\begin{eqnarray}
{\cL}^{(1)}&=&-g_{I\bar{J}}\S^I\bar{\S}^{\bar{J}},\\
{\cL}^{(n+1)}
&=&\cL_{I_1\cdots I_{n+1}\bar{J}_1\cdots \bar{J}_{n+1}}\S^{I_1}\cdots \S^{I_{n+1}}\bar{\S}^{\bar{J}_1}\cdots \bar{\S}^{\bar{J}_{n+1}} \non \\
&=&-{n \over 2(n+1)}{\cL}_{I_1\cdots I_{n-1}\bar{J}_1\cdots\bar{J}_{n}}\S^{I_{n+1}}\bar{\S}^{\bar{J}_{n+1}}{R_{I_{n+1}\bar{J}_{n+1}I_n}}^{L}\S^{I_1}\cdots\S^{I_n}\bar{\S}^{\bar{J_1}}\cdots{\bar{\S}}^{\bar{J}_n}. \label{L1}
\end{eqnarray}
The equation (\ref{L1}) is rewritten as
\begin{eqnarray}
 {\cL}^{(n+1)}={1 \over n+1}\cR_{\S,\bar{\S}}{\cL}^{(n)}, \label{zenka}
\end{eqnarray}
where $\cR_{\S,\bar{\S}}$ is defined as
\begin{eqnarray}
\cR_{\S,{\bar \S}} &=& -\hf \S^K {\bar \S}^{ {\bar L} } \,
R_{K {\bar L} I }{}^J   \big( \F, \bar{\F}  \big)\,\S^I \frac{\pa}{\pa \S^J}.
\label{fd}
\end{eqnarray}
This operator satisfies
\begin{eqnarray}
 \cR_{\S,\bar{\S}}\cL^{(n)}=\bar{\cR}_{\S,\bar{\S}}\cL^{(n)},
\end{eqnarray}
where $\bar{\cR}_{\S,\bar{\S}}$ is the complex conjugate of $\cR_{\S,\bar{\S}}$.
Eq. (\ref{zenka}) yields the explicit form for $\cL$:
\begin{eqnarray}
 \oper{L} = -g_{I\bar J}\Sigma^{I}\frac{e^{\cR_{\Sigma,\bar\Sigma}}-1}{\cR_{\Sigma,\bar\Sigma}} \bar\Sigma^{\bar J}. \label{tbsol}
\end{eqnarray}
On the other hand, acting (\ref{SUSY2}) on $\cL$ leads to the following equation
\begin{eqnarray}
\hf R_{K {\bar J} L }{}^I\, \frac{\pa \cL}{\pa \S^I}\, \S^K \S^L
+ \frac{\pa \cL}{\pa {\bar \S}^{\bar J} }
+g_{I \bar{J}}\, \S^I =0. \label{fd2}
\end{eqnarray}
It is also the condition for invariance under the second SUSY transformation. Indeed, (\ref{tbsol}) satisfies (\ref{fd2}) \footnote{A different form of $\cL$ was obtained in \cite{KuNo}. The solutions in \cite{AKL2} and \cite{KuNo} were shown to be equivalent in \cite{AB}.}.

Next we explain the derivation of the cotangent space part $\cH$ \cite{KuNo,AB}. 
The cotangent bundle action (\ref{act-ctb}) has to be invariant under the following second SUSY transformations \cite{AKL2}
\bea
\d \F^I &=&\hf {\bar D}^2 \big\{ \bar{\ve}_{\dot{\a}} \bar{\q}^{\dot{\a}} \, \S^I  \big(\F, \bar \F, \J , \bar \J \big) \big\} \,, \\
\d \J_I &=&- \hf {\bar D}^2 \Big\{ \bar{\ve}_{\dot{\a}}{\bar{\theta}}^{\dot{\alpha}} \, K_I \big( \F, \bar{\F})  \Big\}
+\hf {\bar D}^2 \Big\{ \bar{\ve}_{\dot{\a}}{\bar{\theta}}^{\dot{\alpha}} \, \G^K_{~IJ} \big( \F, \bar{\F} \big)\,
\S^J  \big(\F, \bar \F, \J , \bar \J \big) \Big\} \,\J_K,
\eea
with 
\bea
\S^I  \big(\F, \bar \F, \J , \bar \J \big) 
= \frac{\pa}{\pa \J_I} \, \cH  \big(\F, \bar \F, \J , \bar \J \big):={\cal H}^I.
\eea
The requirement of invariance under such transformations 
can be shown to be equivalent to the following nonlinear equation \cite{AKL2}:
\begin{eqnarray}
 \cH^Ig_{I\bar{J}}-{1 \over 2}\cH^K\cH^L{R_{K\bar{J}L}}^I\Psi_I=\bar{\Psi}_{\bar{J}}. \label{ce}
\end{eqnarray}
This equation also follows from (\ref{fd2}) using the definition of the $\J$, or it is possible to obtain it by the generalized Legendre transformation \cite{LR1}.
Detailed discussion is given in \cite{AKL2}.

Eq.  (\ref{ce}) implies that
\begin{eqnarray}
 \Psi_I \cH^I - \cH^K\cH^L (R_\Psi)_{KL}=g^{I\bar{J}}\Psi_I\bar{\Psi}_{\bar{J}},\quad (R_\Psi)_{KL}:={1 \over 2}R_{K~L}^{~I~J}\Psi_I\Psi_J. \label{ce3}
\end{eqnarray}
By using the identities
\begin{eqnarray}
 \Psi_I\cH^I=\bar{\Psi}_{\bar{I}}\cH^{\bar{I}}=\sum_{n=1}^\infty n \cH^{(n)},
\end{eqnarray}
(\ref{ce3}) is rewritten as
\begin{eqnarray}
\cH^{(1)}=g^{I\bar{J}}\Psi_I\bar{\Psi}_{\bar{J}},\quad n\cH^{(n)}-\sum_{p=1}^{n-1}\cH^{(p)K}(R_\Psi)_{KL}\cH^{(n-p)L}=0, \quad n\ge 2. \label{ce1}
\end{eqnarray}
In order to solve this, it is useful to define
\begin{eqnarray}
 {\bf R}_{\Psi,\bar{\Psi}}
=\left(
\begin{array}{cc}
0 & (R_\Psi)_I^{~\bar{J}} \\
(R_{\bar{\Psi}})_{\bar{I}}^{~J} & 0
\end{array}
\right)
=\left(
\begin{array}{cc}
0 & {1 \over 2}R_{I}^{~K\bar{J}L}\Psi_K \Psi_L \\
{1 \over 2}R_{\bar{I}}^{~\bar{K}J\bar{L}}\bar{\Psi}_{\bar{K}}\bar{\Psi}_{\bar{L}} & 0
\end{array}
\right) 
\end{eqnarray}
and 
\begin{eqnarray}
&& G^{(2k+2)}=\Psi_Ig^{I\bar{J}}\left(\left(R_{\bar{\Psi}}R_\Psi\right)^k\right)_{\bar{J}}^{~\bar{K}}(R_{\bar{\Psi}})_{\bar{K}}^{~L}\Psi_L=\Psi_{\bar{I}}^\dagger g^{\bar{I}J}\left(\left(R_\Psi R_{\bar{\Psi}}\right)^k\right)_J^{~K}(R_\Psi)_K^{~\bar{L}}\bar{\Psi}_{\bar{L}}, \label{g1} \\
&& G^{(2k+1)}=\Psi_I g^{I\bar{J}}\left(\left(R_{\bar{\Psi}}R_\Psi\right)^k\right)_{\bar{J}}^{~\bar{K}}\bar{\Psi}_{\bar{K}}=\Psi_{\bar{I}}^\dagger g^{\bar{I}J}\left(\left(R_{\Psi}R_{\bar{\Psi}}\right)^k\right)_J^{~K}\Psi_K, \label{g2}
\end{eqnarray}
with $k=0,1,2\cdots$. They satisfy the identities
\begin{eqnarray}
\displaystyle\Psi_I {\partial G^{(n)} \over \partial \Psi_I}&=&\bar{\Psi}_{\bar{I}}{\partial G^{(n)} \over \partial \bar{\Psi}_{\bar{I}}}=nG^{(n)},  \label{dg1} \\
\displaystyle{\partial G^{(2k+2)} \over \partial \Psi_I}
 &=&(2k+2)g^{I\bar{J}}\left(\left(R_{\bar{\Psi}}R_{\Psi}\right)^k\right)_{\bar{J}}^{~\bar{K}}\left(R_{\bar{\Psi}}\right)_{\bar{K}}^{~L}\Psi_L \non \\
 &=&(2k+2)\Psi_Jg^{J\bar{K}}\left(\left(R_{\bar{\Psi}} R_\Psi\right)^k\right)_{\bar{K}}^{~\bar{L}}(R_{\bar{\Psi}})_{\bar{L}}^{~I}, \label{dg2} \\
\displaystyle{\partial G^{(2k+1)} \over \partial \Psi_I}&=&(2k+1)g^{I\bar{J}}\left(\left(R_{\bar{\Psi}}R_{\Psi}\right)^k\right)_{\bar{J}}^{~\bar{K}}\bar{\Psi}_{\bar{K}} \non \\
&=&(2k+1)\Psi^\dagger_{\bar{J}}g^{\bar{J}K}\left(\left(R_{\Psi}R_{\bar{\Psi}}\right)^k\right)_K^{~I}.  \label{dg3}
\end{eqnarray}
Now if we introduce the ansatz
\begin{eqnarray}
 {\cal H}=\sum_{n=1}^{\infty}c_nG^{(n)}, \label{h1}
\end{eqnarray}
where $c_n$ is a constant, we find that the differential equation (\ref{ce1}) turns out to be the algebraic equation
\begin{eqnarray}
nc_n-\sum_{p=1}^{n-1}p(n-p)c_pc_{n-p}=0,\quad c_1=1. \label{ae}
\end{eqnarray}
The equation (\ref{ae}) is universal and independent of the Hermitian symmetric space. Therefore, their solution can be deduced by considering any choice of the Hermitian symmetric space. For instance, the projective complex space ${\bf C}P^1$ can be used. These considerations lead to the solution \cite{KuNo}
\begin{eqnarray}
 {\cal H}(\Phi,\bar{\Phi},\Psi,\bar{\Psi})={1 \over 2}{\bf \Psi}^{\rm T}{\bf g}^{-1}{\cal F}(-{\bf R}_{\Psi,\bar{\Psi}}){\bf \Psi}, \label{ctb}
\end{eqnarray}
where
\begin{eqnarray}
&{\bf \Psi}=\left(
\begin{array}{c}
\Psi_I\\
\bar{\Psi}_{\bar{I}}
\end{array}
\right),\quad
{\bf g}^{-1}=\left(
\begin{array}{cc}
0 & g^{I\bar{J}} \\
g^{\bar{I}J} & 0
\end{array}
\right),& \\
& \displaystyle{\cal F}(x)={1 \over x}\left\{
\sqrt{1+4x}-1-
\ln\left(
1+\sqrt{1+4x} \over 2
\right)
\right\}.&
\end{eqnarray}

\begin{center}
\begin{figure}
 \hspace{2.5cm}
 \includegraphics[width=8cm, bb=0 0 500 272]{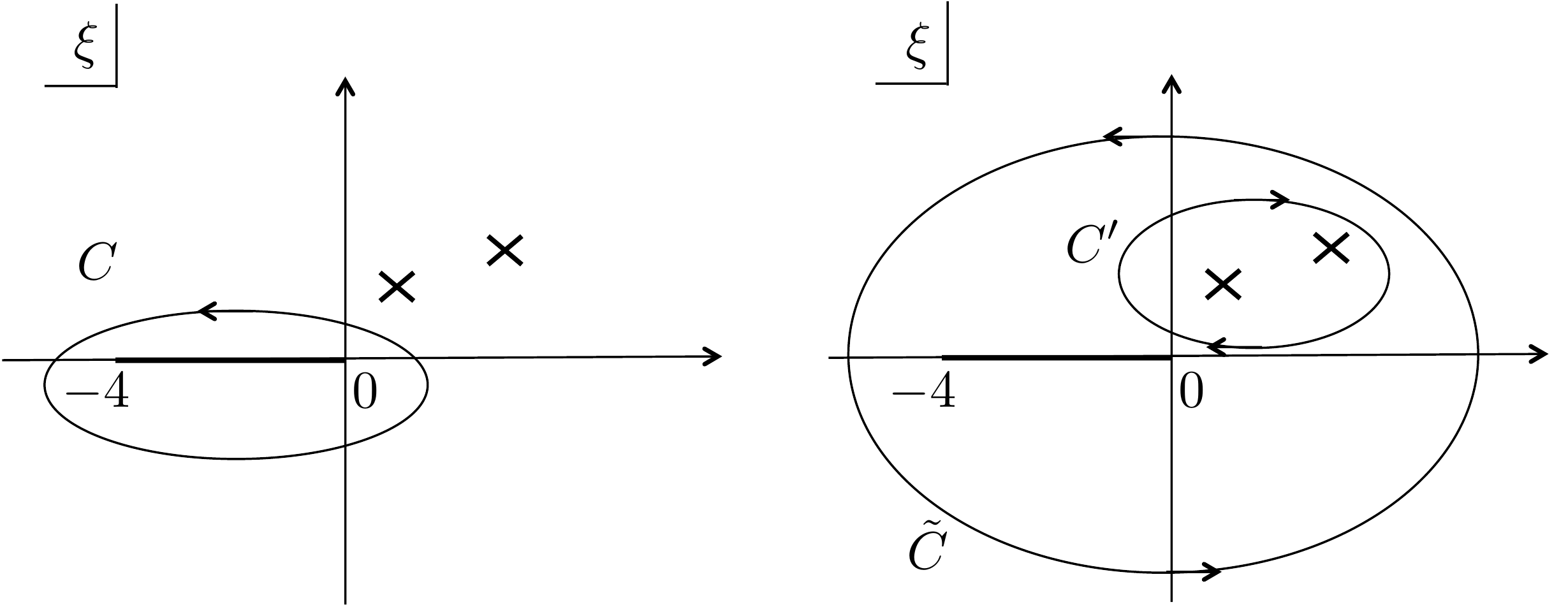}
 \caption{Contour of the integration}\label{contour}
\end{figure}
\end{center}

For our purpose to construct the $\cN=2$ supersymmetric NLSMs on the cotangent bundles  over exceptional Hermitian symmetric spaces, we shall rewrite (\ref{ctb}) into a more convenient form \cite{AB}. First performing the Taylor expansion for (\ref{ctb}) one can see that $\cH$ has the same form as (\ref{h1}) with the coefficient $c_n$ given by
\begin{eqnarray}
 c_n={(-1)^{n-1}{\cal F}^{(n-1)}(0) \over (n-1)!}. \label{cen}
\end{eqnarray}
Second we introduce the differential operators
\begin{eqnarray}
 \cR_{\Psi,\bar{\Psi}}=-(R_{\Psi})_I^{~\bar{J}}\bar{\Psi}_{\bar{J}}{\partial \over \partial \Psi_I},\quad
\bar{\cR}_{\Psi,\bar{\Psi}}=-(R_{\bar{\Psi}})_{\bar{I}}^{~J}\Psi_J{\partial \over \partial \bar{\Psi}_{\bar{I}}}, \label{do}
\end{eqnarray}
which satisfy the following identity
\begin{eqnarray}
 \cR_{\Psi,\bar{\Psi}}\cH^{(n)}=\bar{\cR}_{\Psi,\bar{\Psi}}\cH^{(n)}. \label{hid}
\end{eqnarray}
With the use of (\ref{do}), we can prove that (\ref{g1}) and (\ref{g2}) are compactly written as
\begin{eqnarray}
G^{(n+1)}={(-\cR_{\Psi,\bar{\Psi}})^n \over n!}|\Psi|^2,\quad |\Psi|^2:=g^{I\bar{J}}\Psi_I\bar{\Psi}_{\bar{J}}, \quad n\ge 1. \label{epd}
\end{eqnarray}
Substituting (\ref{epd}) with (\ref{cen}) into (\ref{h1}), we find
\begin{eqnarray}
\cH=\sum_{n=0}^\infty {\cF^{(n)}(0) \over n!}{(\cR_{\Psi,\bar{\Psi}})^n \over n!}|\Psi|^2.
\end{eqnarray}
Making use of the following formula
\begin{eqnarray}
{x^n \over n!}=\oint_C{d\xi \over 2\pi i}{e^{\xi x} \over \xi^{n+1}},
\end{eqnarray}
where the contour $C$ encircles the origin of the complex $\xi$-plane in the counterclockwise direction, we have
\begin{eqnarray}
\cH=\sum_{n=0}^\infty {\cF^{(n)}(0) \over n!}\oint_{C} {d\xi \over 2\pi i}{e^{\xi \cR_{\Psi,\bar{\Psi}}} \over \xi^{n+1}}|\Psi|^2.  \label{f2}
\end{eqnarray}
Here the contour $C$ must be chosen such that the function $e^{\xi \cR_{\Psi,\bar{\Psi}}}|\Psi|^2$ is analytic.
Keeping this in mind, one finds that (\ref{f2}) can be transformed into
\begin{eqnarray}
 {\cal H}=\oint_{C} {d\xi \over 2\pi i}{{\cal F}(1/\xi) \over \xi}e^{\xi \cR_{\Psi,\bar{\Psi}}}|\Psi|^2. \label{fm1}
\end{eqnarray}
The function $\cF(1/\xi)/\xi$ gives a branch cut between $-4$ and $0$. Since the $\cH$ is regular and analytic, the contour $C$ has to be chosen so that it does not cross the branch cut. The resultant contour encircles $\xi=-4, 0$ without crossing the cut and is bounded by poles of the factor $e^{\xi \cR_{\Psi,\bar{\Psi}}}|\Psi|^2$ (see the left figure in Fig. \ref{contour}).  One can transform the contour $C$ as $C^\prime+\tilde{C}$ where $C^\prime$ encircles the poles that may arise from $e^{\xi \cR_{\Psi,\bar{\Psi}}}|\Psi|^2$ in the counterclockwise direction and $\tilde{C}$ encircles those poles together with the branch cut in the clockwise direction (see the right figure in Fig. \ref{contour}). One can check that contribution from the contour $\tilde{C}$ is just a constant by substituting $\xi=Re^{i\theta}$ with $R\rightarrow \infty$ . Therefore, this does not contribute to the K\"ahler metric and can be neglected. We finally have
\begin{eqnarray}
 {\cal H}=-\oint_{-C^\prime}{d\xi \over 2\pi i}{{\cal F}(1/\xi) \over \xi}e^{\xi \cR_{\Psi,\bar{\Psi}}}|\Psi|^2, \label{fm}
\end{eqnarray}
where $-C^\prime$ goes in the counterclockwise direction, yielding the minus sign in front of the integration.

%%%%%%%%%%%%%%%%%%%%%%%%%%%%%%%%%%%%%%%%%%%%%%%%%
%
% Section 3
%
%%%%%%%%%%%%%%%%%%%%%%%%%%%%%%%%%%%%%%%%%%%%%%%%%
\sect{Cotangent bundle over $E_{6(-14)}/SO(10)\times U(1)$}
In this section we derive the $\cN=2$ SUSY NLSM on the cotangent bundle over the non-compact exceptional HSS $E_{6(-14)}/SO(10)\times U(1)$. In the first two subsections we show the $E_{6(-14)}$ algebra and derive the $\cN=1$ SUSY NLSM on $E_{6(-14)}/SO(10)\times U(1)$. There we derive a transformation law of the field parameterizing the space and construct the model so that it is invariant under the transformation. In the subsection \ref{cb-e6}, applying the results in Section 2, we construct the $\cN=2$ SUSY NLSM on the cotangent bundle over $E_{6(-14)}/SO(10)\times U(1)$.

\subsection{Lie algebra of the exceptional group $E_{6(-14)}$}
%%%%%%%%%%%%%%%%%%%%%%%%%%%%%%%%%%%%%%%%%%%%%%%%%
%
% Subsection 1
%
%%%%%%%%%%%%%%%%%%%%%%%%%%%%%%%%%%%%%%%%%%%%%%%%%
The non-compact group $E_{6(-14)}$ has the maximal compact subgroup
$SO(10) \times U(1)$.
A way to write down the Lie algebra of $E_{6(-14)}$ is to use
the relation between $E_{6(-14)}$ and the pair $(E_{6(-14)}, SO(10) \times U(1))$.
The $E_{6(-14)}$ algebra is given by the generators
$T_{AB} (A, B = 1, \cdots 10)$ of $SO(10)$, $T$ of $U(1)$,
and $SO(10)$ Majorana spinors $(E_{\alpha}, \bar{E}^{\alpha}) (\alpha = 1, \cdots, 16)$ with $\bar{E}^{\alpha} = - (E_{\alpha})^{\dagger}$
satisfying the following commutation relations:
\begin{eqnarray}
 &[T_{AB}, T_{CD}]=\delta_{BC}T_{AD}+\delta_{AD}T_{BC}-\delta_{AC}T_{BD}-\delta_{BD}T_{AC},& \label{SO(10)}\\
 & [E_\alpha, T_{AB}]=(\sigma_{AB})_\alpha^{~\beta}E_\beta, \quad
  [\bar{E}^\alpha, T_{AB}]=(\bar{\sigma}_{AB})^\alpha_{~\beta}\bar{E}^\beta, 
& \\
 & \displaystyle [T,E_\alpha]={\sqrt{3} \over 2}iE_\alpha,\quad [T, \bar{E}^\alpha]=-{\sqrt{3} \over 2}i\bar{E}^\alpha, & \\
 & [E_\alpha,E_\beta]=[\bar{E}^\alpha, \bar{E}^\beta]=0, & \\
 & \displaystyle [E_\alpha, \bar{E}^\beta]={1 \over 2}(\sigma_{AB})_\alpha^{~ \beta}T_{AB}-{\sqrt{3} \over 2}i\delta_\alpha^{~\beta}T,  & \label{E}
\end{eqnarray}
where $\sigma_{AB}$ and $\bar{\sigma}_{AB}$ are the generators of the Weyl 
representation of $SO(10)$ group given in (\ref{sog}).
Here, by making replacement $E_{\alpha} \to iE_{\alpha}, \bar{E}^{\alpha}\to i\bar{E}^{\alpha}$ in the relations (\ref{SO(10)}) -- (\ref{E}),
the corresponding algebra
gives the Lie algebra of the compact $E_{6}$ \cite{He},
which is given in \cite{AcAoHo,DV,ItKuKu}.

%%%%%%%%%%%%%%%%%%%%%%%%%%%%%%%%%%%%%%%%%%%%%%%%%
%
% Subsection 2
%
%%%%%%%%%%%%%%%%%%%%%%%%%%%%%%%%%%%%%%%%%%%%%%%%%
\subsection{K\"ahler potential}
Now we construct the ${\cal N}=1$ SUSY NLSM whose K\"ahler potential is
of 
$\cM=E_{6(-14)}/SO(10)\times U(1)$.
The fields in the NLSM can be interpreted as the Nambu-Goldstone boson when the symmetry $E_{6(-14)}$ is broken down to $SO(10)\times U(1)$. The unbroken generators are $E_\alpha$ and its complex conjugate $\bar{E}^\alpha$. Corresponding representation is of the 16 complex Weyl spinor of $SO(10)$. We write it with Greek index appeared in the algebra as
\begin{eqnarray}
 \Phi^I \rightarrow \Phi_\alpha,\quad \bar{\Phi}^I \rightarrow \bar{\Phi}^\alpha.
\end{eqnarray}
The K\"ahler potential is constructed so that it is invariant under the  transformation of $\Phi_\a$. The transformation law for $\Phi_\a$ is read from
the following commutation relations\footnote{The corresponding transformation
for the compact group $E_{6}$ is found in \cite{AcAoHo,DV}.}:
\begin{eqnarray}
 & [\Phi_\alpha, T_{AB}]=(\sigma_{AB})_\alpha^{~\beta}\Phi_\beta,\quad 
 [\bar{\Phi}^{\alpha}, T_{AB}]=(\bar{\sigma}_{AB})^{\alpha}_{~\beta}\bar{\Phi}^\beta,& \label{ec1} \\
 & \displaystyle [T, \Phi_\alpha]={\sqrt{3} \over 2}i\Phi_\alpha, \quad [T, \bar{\Phi}^\alpha]=-{\sqrt{3} \over 2}i\bar{\Phi}^\alpha, & \label{ec2} \\
 & \displaystyle [E_\alpha, \Phi_\beta]={1 \over 4}\Sigma_{\alpha\beta}^{\gamma\delta}\Phi_\gamma\Phi_\delta, 
   \quad [E_\alpha,\bar{\Phi}^\beta]=\delta_\alpha^{~\beta}, & \label{ec3} \\
 & \displaystyle [\bar{E}^\alpha, \Phi_\beta]=\delta_\beta^{~\alpha},\quad 
   [\bar{E}^\alpha, \bar{\Phi}^\beta]={1 \over 4}\Sigma_{\gamma\delta}^{\alpha\beta}\bar{\Phi}^\gamma\bar{\Phi}^\delta,\label{ec4}
\end{eqnarray}
where 
\begin{eqnarray}
 \Sigma_{\alpha\beta}^{\gamma\delta}=\sum_{A,B}(\sigma_{AB})_\alpha^{~\gamma}(\sigma_{AB})_\beta^{~\delta}-{3 \over 2}\delta_\alpha^{~\gamma}\delta_\beta^{~\delta}
\end{eqnarray}
satisfying the properties:
\begin{eqnarray}
& \Sigma_{\alpha\beta}^{\gamma\delta}=\Sigma_{\beta\alpha}^{\gamma\delta}
=\Sigma_{\alpha\beta}^{\delta\gamma}, & \label{p1} \\
& \Sigma_{\rho\sigma}^{\lambda(\alpha}\Sigma_{\tau\lambda}^{\beta\gamma)}
=\Sigma_{\sigma\tau}^{\lambda(\alpha}\Sigma_{\rho\lambda}^{\beta\gamma)}
=\Sigma_{\tau\rho}^{\lambda(\alpha}\Sigma_{\sigma\lambda}^{\beta\gamma)}. & \label{p2}
\end{eqnarray}
The closure of the $E_{6(-14)}$ algebra on $\Phi_\alpha$ can be checked by using the Jacobi identities involving two generators of $E_{6(-14)}$ algebra and one $\Phi_\alpha$. For instance, one can check that the following non-trivial identity is satisfied
\begin{eqnarray}
 [E_\alpha,[{E}_\beta,\Phi_{\gamma}]]+(\mbox{cyclic~permutations})=0, 
\end{eqnarray}
by using (\ref{p2}).
The commutators (\ref{ec3})--(\ref{ec4})
lead to the infinitesimal transformation law for $\Phi_\alpha$ as
\begin{eqnarray}
 \delta\Phi_\alpha&=&\bar{\epsilon}^\beta[E_\beta, \Phi_\alpha]+\epsilon_\beta[\bar{E}^\beta,\Phi_\alpha] \nonumber\\
 &=&\epsilon_\alpha+{1 \over 4}\bar{\epsilon}^\beta\Sigma_{\alpha\beta}^{\gamma\delta}\Phi_\gamma\Phi_\delta, \label{trans1}
\end{eqnarray}
where $\epsilon_\alpha$ and $\bar{\epsilon}^\beta$ are transformation parameters.
 
For convenience, we embed the transformation law for the 16-component  spinor (\ref{trans1}) to one for the 32-component spinor. In other words, we embed the irreducible Weyl spinor $\Phi_\alpha$ into the reducible Dirac spinor $\phi_a$ of $SO(10)$:
\begin{eqnarray}
\Phi_\alpha \rightarrow \phi_{a}\equiv
 \left(
 \begin{array}{c}
  \Phi_\alpha \\
  {\bf 0}
 \end{array}
\right)
=(P^+\phi)_a, \quad a=1,\cdots 32, \label{w1}
\end{eqnarray}
where $P^+$ is the projection operator defined by (\ref{app.eq.proj}).
Similarly, we embed the transformation parameter $\epsilon_\alpha$ into the 32-component spinor $\varepsilon_a$
\begin{eqnarray}
 \epsilon_\alpha\rightarrow\varepsilon_a\equiv  
\left(
 \begin{array}{c}
  \epsilon_\alpha \\
  {\bf 0}
 \end{array}
\right)
=(P^+\varepsilon)_a. \label{w2}
\end{eqnarray}
The generator $\sigma_{AB}$ in the Weyl representation forms the Dirac representation as
\begin{eqnarray}
 (\sigma_{AB})_\alpha^{~\beta}\rightarrow \Gamma_{AB}^+\equiv P^+\Gamma_{AB}={1 \over 2}P^+[\Gamma_A,\Gamma_B], \label{w3}
\end{eqnarray}
where $\Gamma_{A}$ is a $32\times 32$ gamma matrix defined by (\ref{gamma}). 
With the use of (\ref{w1})--(\ref{w3}), (\ref{trans1}) is rewritten by
\begin{eqnarray}
 \delta\phi_a=\varepsilon_a+{1 \over 16}\sum_{A,B}(\bar{\varepsilon}^{\rm T}\Gamma_{AB}^+\phi)(\Gamma_{AB}^+\phi)_a-{3 \over 8}(\bar{\varepsilon}^{\rm T}\phi)(P^+\phi)_a. \label{trans2}
\end{eqnarray}
Defining the charge conjugation
\begin{eqnarray}
 \phi_c=\phi^{\rm T}C,\quad \bar{\varepsilon}_c=C^{-1}\bar{\varepsilon},
\end{eqnarray}
where $C$ is the charge conjugation matrix given by (\ref{ccm}), (\ref{trans2}) turns out to be
\begin{eqnarray}
\delta\phi_a=\varepsilon_a-{1 \over 16}\sum_{A,B}(\phi_{c}\Gamma_{AB}^-\bar{\varepsilon}_c)(\Gamma_{AB}^+\phi)_a-{3 \over 8}(\phi_cP^-\bar{\varepsilon}_c)(P^+\phi)_a, \label{trans3}
\end{eqnarray}
with the projection operator $P^-$ defined by (\ref{app.eq.proj}). Here we have used (\ref{app.eq.gamma}).
By using the Fierz identity we exchange $\bar{\varepsilon}^a$ and $\phi_a$ in (\ref{trans3}) and obtain
\begin{eqnarray}
\delta\phi_a=\varepsilon_a+{1 \over 4}\sum_{A}(\phi_{c}\Gamma_{A}^+\phi)(\Gamma_{A}^-\bar{\varepsilon}_c)_a-(\bar{\varepsilon}^{\rm T}\phi)(P^+\phi)_a, \label{trans4}
\end{eqnarray}
where $\Gamma_A^{\pm}\equiv \Gamma_AP^\pm$.
The detailed calculation to obtain (\ref{trans4}) from (\ref{trans3}) is summarized in Appendix \ref{calc}.

Let us derive the K\"ahler potential invariant under the nonlinear transformation law (\ref{trans4}). In order to do that, first we consider the $SO(10)\times U(1)$ invariants:
\begin{eqnarray}
 I_1&=&\bar{\phi}^{\rm T}\phi,  \\
 I_2&=&{1 \over 8}\sum_{A}(\phi_c\Gamma_A^+\phi)^\dagger(\phi_c\Gamma_A^+\phi).
\end{eqnarray}
They transform under (\ref{trans4}) as
\begin{eqnarray}
 \delta I_1&=&(1-I_1)(\bar{\varepsilon}^{\rm T}P^+\phi)+{1 \over 4}\sum_A(\bar{\phi}^{\rm T}\Gamma_A^-\bar{\varepsilon}_c)(\phi_c\Gamma_A^+\phi)+{\rm c.c,} \label{inv1}\\
 \delta I_2&=&{1 \over 4}\sum_A(\bar{\phi}^{\rm T}\Gamma_A^-\bar{\varepsilon}_c)(\phi_c\Gamma_A^+\phi)-(\bar{\varepsilon}^{\rm T}P^+\phi)I_2+{\rm c.c.} \label{inv2}
\end{eqnarray}
In deriving (\ref{inv2}) we have used the transformation law for the 10-dimensional representation of the $SO(10)\times U(1)$ from the product of spinors $16\times 16$,
$\phi_c\Gamma_A^+\phi$. It is given by
\begin{eqnarray}
 \delta(\phi_c\Gamma_A^+\phi)=2(\phi_c\Gamma_A^+\epsilon)
  -(\phi_c\Gamma_A^+\phi)(\bar{\varepsilon}^{\rm T}P^+\phi). \label{trans5}
\end{eqnarray}
It can be checked by starting with
\begin{eqnarray}
 \delta(\phi_c\Gamma_A^+\phi)=2(\phi_c\Gamma_A^+\varepsilon)-2(\phi_c\Gamma_A^+\phi)(\bar{\varepsilon}P^+\phi)+{1 \over 2}\sum_B(\phi_c\Gamma_B^{+}\phi)(\bar{\varepsilon}^{\rm T}\Gamma_B^+\Gamma_A^-\phi_c).
\end{eqnarray}
Applying (\ref{com}) to the last term in the right-hand side and using (\ref{i4}), we obtain (\ref{trans5}).

Using the relations (\ref{inv1}) and (\ref{inv2}) one can check that the following function
\begin{eqnarray}
 K(\phi,\bar{\phi})=-\ln(1-I_1+I_2) \label{kahler}
\end{eqnarray}
transforms as
\begin{eqnarray}
 \delta K=(\bar{\varepsilon^{\rm T}}P^+\phi)+{\rm c.c.} \label{kahlert}
\end{eqnarray}
The right-hand side consists of the holomorphic and anti-holomorphic parts with respect to $\phi$.
It means that (\ref{kahler}) is invariant under (\ref{trans2}) up to the K\"ahler transformation (\ref{kahlert}).
Therefore we conclude that (\ref{kahler}) is the K\"ahler potential of $E_{6(-14)}/SO(10)\times U(1)$.

Finally for later use we shall go back to the Weyl representation. 
Using the representation of the gamma matrix (\ref{gamma}),
the K\"ahler potential is written by
\begin{eqnarray}
 K(\Phi, \bar{\Phi})=-\ln\left(1-\bar{\Phi}^{\rm T}\Phi+{1 \over 8}\sum_A(\bar{\Phi}^{\rm T}\sigma_A\cC^\dagger\bar{\Phi})(\Phi^{\rm T}\cC\sigma_A^\dagger\Phi)\right), \label{K}
\end{eqnarray}
where $\cC$ is the charge conjugation matrix in the Weyl representation defined by (\ref{ccm}).
The former sign is not determined by using the transformation law (\ref{trans4}) but 
it is chosen to ensure positivity of the metric. The K\"ahler potential (\ref{K}) 
is very similar to one of the compact HSS $E_6/SO(10)\times U(1)$ given by \cite{DV, ItKuKu, HiNi}\footnote{A different form the K\"ahler potential is given in \cite{AcAoHo}.}. Only the differences are the sign in front of the logarithm and in the second term inside the logarithm.

%%%%%%%%%%%%%%%%%%%%%%%%%%%%%%%%%%%%%%%%%%%%%%%%%
%
% Subsection 3
%
%%%%%%%%%%%%%%%%%%%%%%%%%%%%%%%%%%%%%%%%%%%%%%%%%
\subsection{Cotangent bundle}\label{cb-e6}
Let us derive the cotangent bundle over $E_{6(-14)}/SO(10)\times U(1)$ with use of the formula (\ref{fm}).
Since we are considering the symmetric space, it is sufficient to consider (\ref{fm}) at $\Phi=\bar{\Phi}=0$. 
We shall introduce the notation of the cotangent vectors:
\begin{eqnarray}
\Psi^I \rightarrow \Psi^\alpha,\quad \bar{\Psi}^{\bar{I}}\rightarrow \bar{\Psi}_\a.
\end{eqnarray}
The metric and Riemann tensor at $\F=0$ are derived from the K\"ahler potential (\ref{K}) as
\begin{eqnarray}
g^\a_{~\b}{\Big |}_{\F=\bar{\F}=0}&=&
 {\partial^2 K \over \partial_\alpha \Phi \partial^\beta\bar{\Phi}}{\Big |}_{\F=\bar{\F}=0}=
\delta^\a_{~\b},\label{m1} \\
R^{\a~\g}_{~\b~\d}{\Big |}_{\F=\bar{\F}=0}&=&\partial^\g\partial_\d g^{\a}_{~\b}
 -(g^{-1})^\l_{~\k}\partial^\k g^\a_{~\b}\partial_\l g^\g_{~\d}{\Big
 |}_{\F=\bar{\F}=0} \nonumber \\
 &=&\d_\d^{~\g}\d_\b^{~\a}
 -\frac{1}{ 2}\sum_A(\s_A\cC^\dagger)_{\b\d}(\cC\s_A^\dagger)^{\a\g}+\d_\b^{~\g}\d_\d^{~\a}, \label{r1}
\end{eqnarray}
where $(g^{-1})^\b_{~\a}=(g^\a_{~\b})^{-1}$ is the inverse metric of $g^\a_{~\b}$, namely $g^\a_{~\g}(g^{-1})^\gamma_{~\beta}=\delta^\a_{~\b}$. 
We calculate the differential operator (\ref{do}) in (\ref{fm}), which is written as
\begin{eqnarray}
\cR_{\Psi,\bar{\Psi}}{\Big |}_{\Phi=\bar{\Phi}=0}=-|\psi^2|\psi^\a{\partial \over \partial \psi^\a}+{1 \over 4}\sum_A(\psi\sigma_A\cC\psi)(\bar{\psi}\cC\sigma_A^\dagger)^\a{\partial \over \partial \psi^\a},  \label{do2}
\end{eqnarray}
where $\psi^\a$ is the cotangent vector at $\Phi=\bar{\Phi}=0$.
Defining the $SO(10)\times U(1)$ invariants
\begin{eqnarray}
 &&x:=\psi^\a\bar{\psi}_\a, \\
 &&y:=(\psi\sigma_A\cC\psi)(\bar{\psi}\cC\sigma_A^\dagger\bar{\psi}),
\end{eqnarray}
we find that (\ref{do2}) is rewritten by
\begin{eqnarray}
 \cR_{\Psi,\bar{\Psi}}{\Big |}_{\Phi=\bar{\Phi}=0}=-xD+{1 \over 4}y\partial_x,\quad
  D:=x{\partial \over \partial x}+y{\partial \over \partial y}.
\end{eqnarray}
Here we have used the identity 
\begin{eqnarray}
 \sum_A(\sigma_A^\dagger \psi)(\S\cC\psi^\dagger_A\psi)=0,
\end{eqnarray} 
and $\{\sigma_A,\sigma_B\}=2\delta_{AB}$.
Using the Baker-Campbell-Hausdorff formula, we find that (\ref{fm1}) is
\begin{eqnarray}
\cH&=&-\oint_{-C^\prime}{\cF(1/\xi) \over \xi}e^{-\xi xD}e^{{1 \over 4}\xi y {\partial \over \partial x}}e^{-{1 \over 8}\xi^2yD}x \non \\
 &=&-\oint_{-C^\prime}{\cF(1/\xi) \over \xi}{\partial \over \partial \xi}\ln\left(1+\xi x+{\xi^2 y \over 8}\right). \label{int1}
\end{eqnarray}
Recalling that the contour $-C^\prime$ encircles the poles of $e^{\xi\cR_{\Psi,\bar{\Psi}}}|\Psi|^2$, one sees that the roots of the equation
\begin{eqnarray}
1+\xi x+{\xi^2 y \over 8}=0, \label{root0}
\end{eqnarray}
contribute to the integral (\ref{int1}). 
Considering this, (\ref{int1}) is calculated by the Residue theorem
\begin{eqnarray}
 \cH = -\left({\cF(1/\xi_+) \over \xi_+}+{\cF(1/\xi_-) \over \xi_-}\right), \label{fr1}
\end{eqnarray}
where $\xi_{\pm}$ are the solutions of (\ref{root0}) given by
\be
\xi_{\pm}={-4x\pm 2\sqrt{4x^2-2y} \over y}.
\ee

Alternatively, one can obtain the cotangent bundle action from the tangent bundle action with the use of (\ref{tbsol}) and the generalized Legendre transformation \cite{LR1}. Detailed calculations are found in Appendix \ref{GLT}. The result at an arbitrary point of the base manifold is also given there.
%%%%%%%%%%%%%%%%%%%%%%%%%%%%%%%%%%%%%%%%%%%%%%%%%
%
% Section 4
%
%%%%%%%%%%%%%%%%%%%%%%%%%%%%%%%%%%%%%%%%%%%%%%%%%
\sect{Cotangent bundle over $E_{7(-25)}/E_6\times U(1)$}
In this section, we derive the $\cN=2$ SUSY NLSM on the cotangent bundle over $E_{7(-25)}/E_6\times U(1)$.
We perform the same procedure as in Section 3: We first derive the $\cN=1$ SUSY NLSM on $E_{7(-25)}/E_6\times U(1)$ which is found by checking invariance under the nonlinear transformation of the field parameterizing $E_{7(-25)}/E_6\times U(1)$. Second using this result with the formula in Section 2, we construct the $\cN=2$ SUSY model on the cotangent bundle over $E_{7(-25)}/E_6\times U(1)$.

%%%%%%%%%%%%%%%%%%%%%%%%%%%%%%%%%%%%%%%%%%%%%%%%%
%
% Subsection 4.1
%
%%%%%%%%%%%%%%%%%%%%%%%%%%%%%%%%%%%%%%%%%%%%%%%%%
\subsection{Lie algebra of the exceptional group $E_{7(-25)}$}
The non-compact group $E_{7(-25)}$ has the subgroup $E_6\times U(1)$. 
The Lie algebra of $E_{7(-25)}$ is written by the generators of this subgroup that are 
$E_6$ generator $T_A(A=1,\dots 78)$, a $U(1)$ generator $T$, 
$E^i(i=1,\dots 27)$ belonging to the $E_6$ fundamental representation and their conjugates $\bar{E}_i=-(E^i)^\dagger$.
The commutation relations are given as
\begin{eqnarray}
& [T_A, T_B]=-f_{AB}^{~~C}T_C, \quad [T, T_A]=0, \label{a1} &\\
& [T_A, E^i]=i\rho(T_A)^i_{~j}E^j,\quad [T_A,\bar{E}_i]=-i\bar{E}_j\rho(T_A)^j_{~i}, & \\
& \displaystyle [T, E^i]=i\sqrt{2 \over 3}E^i,\quad [T,\bar{E}_i]=-i\sqrt{2 \over 3}\bar{E}_i,& \\
& [E^i, E^j]=[\bar{E}_i,\bar{E}_j]=0, &\\
& \displaystyle [E^i, \bar{E}_j]=i\sum_A\rho(T_A)^i_{~j}T_A+i\sqrt{2 \over 3}\delta^i_{~j}T,\label{a2} &
\end{eqnarray}
where $f_{AB}^{~~C}$ are structure constants of $E_6$ and $\rho(T_A)$ is a fundamental representation matrix of $E_6$ which satisfies \cite{KeVa,Cv}
\begin{eqnarray}
 \sum_A\rho(T_A)^i_{~j}\rho(T_A)^l_{~k}={1 \over 6}\delta^l_{k}\delta^i_{j}+{1 \over 2}\delta^i_{k}\delta^l_{j}-{1 \over 2}\Gamma^{ilp}\Gamma_{pjk}. \label{g6}
\end{eqnarray}
Here $\Gamma^{ijk}$ is the invariant tensor of $E_6$. This is symmetric with respect to the indices $(i,j,k)$ and the complex
conjugate is defined as $(\Gamma^{ijk})^\dagger=\Gamma_{ijk}$. This satisfies the following identity
\begin{eqnarray}
& \Gamma_{ijk}\Gamma^{ljk}=10\delta_i^l, &
\end{eqnarray}
and the Springer relation \cite{Sp}
\begin{eqnarray}
\Gamma^{ijk}\Gamma_{jl(m}\Gamma_{pq)k}=\delta^i_{(l}\Gamma_{mpq)}. \label{sp}
\end{eqnarray}

Note that replacements $E^i\rightarrow iE^i$ and $\bar{E}_i\rightarrow i\bar{E}_i$ in (\ref{a1})--(\ref{a2}) give the compact Lie algebra of $E_7$ \cite{HiNi}.

%%%%%%%%%%%%%%%%%%%%%%%%%%%%%%%%%%%%%%%%%%%%%%%%%
%
% Subsection 4.2
%
%%%%%%%%%%%%%%%%%%%%%%%%%%%%%%%%%%%%%%%%%%%%%%%%%
\subsection{K\"ahler potential}
In this subsection, we construct the $\cN=1$ SUSY NLSM whose K\"ahler potential is of $E_{7(-25)}/E_6\times U(1)$. 
The K\"ahler potential is written by the coordinates which are representation of $E^i$ and $\bar{E}_i$. We shall write them as
\begin{eqnarray}
 \Phi^I \rightarrow \phi^i,\quad \bar{\Phi}^{\bar{I}}\rightarrow \bar{\phi}_i.
\end{eqnarray}
The transformation law for $\phi$ is obtained from the commutation relations:
\begin{eqnarray}
 & \displaystyle [E^i, \phi^j]=\phi^i\phi^j-{1 \over 2}\Gamma^{ijk}\Gamma_{klm}\phi^l\phi^m, & \label{cr1}\\
 & [E^i,\bar{\phi}_j]=\delta^i_j,\quad [\bar{E}_i,\phi^j]=\delta_i^j, \label{cr2} & \\
 & \displaystyle[\bar{E}_i,\bar{\phi}_j]=\bar{\phi}_i\bar{\phi}_j-{1 \over 2}\Gamma_{ijk}\Gamma^{klm}\bar{\phi}_l\bar{\phi}_m, \label{cr3} & \\
 & [T_A, \phi^i]=-2i\rho(T_A)^i_{~j}\phi^j,\quad [T_A, \bar{\phi}_i]=2i\bar{\phi}_j\rho(T_A)^j_{~i}, \label{cr4} &\\
 & \displaystyle [T,\phi^i]=-i\sqrt{2 \over 3}\phi^i,\quad [T, \bar{\phi}_i]=i\sqrt{2 \over 3}\bar{\phi}_i. \label{cr5}&
\end{eqnarray}
One can see the closure of this algebra by checking the Jacobi identity. A nontrivial one is, for example,
\begin{eqnarray}
 [E^i,[\bar{E}_j,\bar{\phi}_k]]+({\rm cyclic~permutation})=0,
\end{eqnarray}
which can be proved with the aid of (\ref{g6}). One can also prove another nontrivial identity 
\begin{eqnarray}
 [E^i,[E^j,\phi^k]]+({\rm cyclic~permutation})=0,
\end{eqnarray}
by utilizing (\ref{sp}).

The commutation relations (\ref{cr1})--(\ref{cr3}) lead to the infinitesimal transformation law for $\phi^i$:
\begin{eqnarray}
 \delta\phi^i&=&\epsilon^j[\bar{E}_j, \phi^i]+\bar{\epsilon}_j[E^j,\phi^i] \non \\
 &=&\epsilon^i-(\bar{\epsilon}_j\phi^j)\phi^i+{1 \over 2}\Gamma^{ijk}\bar{\epsilon}_j\Gamma_{klm}\phi^l\phi^m, \label{tf2}
\end{eqnarray}
where $\epsilon^i$ is the complex transformation parameter.

Now we look for the function being invariant under (\ref{tf2}). To this end, we introduce the $E_6\times U(1)$ invariants:
\begin{eqnarray}
&&I_1=\bar{\phi}_i\phi^i, \\
&&I_2=(\Gamma_{ijk}\phi^j\phi^k)(\Gamma^{ilm}\bar{\phi}_l\bar{\phi}_m), \\
&&I_3={1 \over 9}(\Gamma_{ijk}\phi^i\phi^j\phi^k)(\Gamma^{lmn}\bar{\phi}_l\bar{\phi}_m\bar{\phi}_n).
\end{eqnarray}
They transform under (\ref{tf2}) as
\begin{eqnarray}
&&\delta I_1=(1-I_1)(\bar{\epsilon}_i\phi^i)+{1 \over 2}(\Gamma_{ijk}\phi^j\epsilon^k)(\Gamma^{ilm}\bar{\phi}_l\bar{\phi}_m)+{\rm c.c.} \label{e7i1} \\
&&\delta I_2=2(\Gamma_{ijk}\phi^j\phi^k)(\Gamma^{ilm}\bar{\phi}_l\bar{\phi}_m)-(\bar{\epsilon}_i\phi^i)I_2+{1 \over 3}(\Gamma_{ijk}\phi^i\phi^j\phi^k)(\Gamma^{lmn}\bar{\phi}_l\bar{\phi}_m\bar{\phi}_n)+{\rm c.c.} \label{e7i2} \\
&&\delta I_3=-(\bar{\epsilon}_i\phi^i)I_3+{1 \over 3}(\Gamma_{ijk}\phi^i\phi^j\phi^k)(\Gamma^{lmn}\bar{\epsilon}_l\bar{\phi}_m\bar{\phi}_n)+{\rm c.c.} \label{e7i3}
\end{eqnarray}
By using (\ref{e7i1})--(\ref{e7i3}), one can check that the function 
\begin{eqnarray}
 K=-\ln\left(1-I_1+{1 \over 4}I_2-{1 \over 4}I_3\right) \label{ke7}
\end{eqnarray}
transforms under the infinitesimal transformation (\ref{tf2}) as
\begin{eqnarray}
 \delta K=(\bar{\epsilon}_i\phi^i)+{\rm c.c.} \label{k2}
\end{eqnarray}
This shows that (\ref{ke7}) is invariant under (\ref{tf2}) up to the K\"ahler transformation (\ref{k2}). Thus we conclude that (\ref{ke7}) is the K\"ahler potential of $E_{7(-25)}/E_6\times U(1)$. As in the case for $E_{6(-14)}/SO(10)\times U(1)$, the sign in front of the logarithm in (\ref{ke7}) cannot be determined by the invariance under (\ref{tf2}). It is just chosen so that positivity of the metric is ensured.

%%%%%%%%%%%%%%%%%%%%%%%%%%%%%%%%%%%%%%%%%%%%%%%%%
%
% Subsection 4.4
%
%%%%%%%%%%%%%%%%%%%%%%%%%%%%%%%%%%%%%%%%%%%%%%%%%
\subsection{Cotangent bundle}
We derive the cotangent bundle action of $E_{7(-25)}/E_6\times U(1)$ by using (\ref{fm}).
First we introduce the cotangent vectors:
\begin{eqnarray}
 \Psi\rightarrow \Psi_i, \quad \bar{\Psi}^{\bar{I}}\rightarrow \bar{\Psi}^i
\end{eqnarray}
Since we are considering the symmetric space, we set $\phi=\bar{\phi}=0$ in calculations.
The metric and the Riemann tensor at $\phi=\bar{\phi}=0$ are derived from (\ref{ke7}):
\begin{eqnarray}
&& g_i^{~j}{\Big |}_{\phi=\bar{\phi}=0}={\partial^2 K \over \partial \phi^i \partial \bar{\phi}_j}{\Big |}_{\phi=\bar{\phi}=0}=\delta_i^{~j}, \label{m2}\\
&& R_{i~k}^{~j~l}{\Big |}_{\phi=\bar{\phi}=0}=\delta_i^{~l}\delta_k^{~j}-\Gamma_{mik}\Gamma^{mjl}+\delta_k^{~l}\delta_i^{~j}. \label{r2}
\end{eqnarray}
Then the differential operator (\ref{do}) at $\phi=\bar{\phi}=0$ is obtained as
\begin{eqnarray}
 \cR_{\Psi,\bar{\Psi}}{\Big |}_{\phi=\bar{\phi}=0}=-|\psi|^2\psi_i{\partial \over \partial \psi_i}+{1 \over 2}(\Gamma^{mik}\psi_i\psi_k)\Gamma_{mlj}\psi^j{\partial \over \partial \psi_l}, \quad |\psi|^2:=\psi_i\bar{\psi}^i, \label{do3}
\end{eqnarray}
where $\psi$ and $\bar{\psi}$ are coordinates of the cotangent space at $\phi=\bar{\phi}=0$.
If we define the $E_6\times U(1)$ invariants in terms of the cotangent vector
\begin{eqnarray}
&& x:=\psi_i\bar{\psi}^i, \\
&& y:=(\Gamma^{ijk}\psi_j\psi_k)(\Gamma_{ilm}\bar{\psi}^l\bar{\psi}^m), \\
&& z:=(\Gamma^{ijk}\psi_i\psi_j\psi_k)(\Gamma_{lmn}\bar{\psi}^l\bar{\psi}^m\bar{\psi}^n),
\end{eqnarray}
the differential operator (\ref{do3}) is rewritten as
\begin{eqnarray}
 \cR_{\Psi,\bar{\Psi}}{\Big |}_{\phi=\bar{\phi}=0}=xD-{1 \over 2}y{\partial \over \partial x}-{1 \over 3}z{\partial \over \partial y},\quad
 D:=x{\partial \over \partial x}+y{\partial \over \partial y}+z{\partial \over \partial z}.
\end{eqnarray}
The factor $e^{\xi \cR_{\Psi,\bar{\Psi}}}x$ in (\ref{fm}) is calculated by using the Baker-Campbell-Hausdorff formula
\begin{eqnarray}
 e^{\xi \cR_{\Psi,\bar{\Psi}}}x{\Big |}_{\phi=\bar{\phi}=0}
 &=&e^{-\xi xD}e^{{1 \over 2}\xi y {\partial \over \partial x}}e^{{1 \over 3}\xi z {\partial \over \partial y}}e^{{1 \over 12}\xi^2 z {\partial \over \partial x}}e^{-{1 \over 4}\xi^2 y D}e^{{1 \over 18}\xi^3z D}x \non \\
&=& {\partial \over \partial \xi}\ln\left(1+\xi x+{1 \over 4}\xi^2 y+{1 \over 36}\xi^3 z\right).
\end{eqnarray}
The poles arising from this factor contribute to the integral in (\ref{fm}), which are obtained from the equation
\be
1+\xi x+{1 \over 4}\xi^2 y+{1 \over 36}\xi^3 z={z \over 36}(\xi-\xi_1)(\xi-\xi_2)(\xi-\xi_3)=0.
\ee
The poles are obtained as
\begin{eqnarray}
 \xi_1&=& -{3y \over z}-{2^{1/3}(-81y^2+108xz) \over 3z(-1458y^3+2916xyz-972z^2+A)^{1/3}} \non \\
         &&~\quad +{(-1458y^3+2916xyz-972z^2+A)^{1/3} \over 3 \cdot 2^{1/3}z}, \\\label{e7sol3}
 \xi_2&=& -{3 y \over z} + {(1 + i\sqrt{3}) (-81 y^2 + 108 x z) \over 3\cdot 2^{2/3}
     z (-1458 y^3 + 2916 xyz - 972 z^2 + A)^{1/3}} \non \\
&& - {1 - i\sqrt{3} \over 6\cdot 2^{1/3} z} (-1458 y^3 + 2916 x y z - 972 z^2 
    +  A)^{1/3},\\\label{e7sol4}
 \xi_3&=& -{3 y \over z} + {(1 - i\sqrt{3}) (-81 y^2 + 108 x z) \over 3\cdot 2^{2/3}
     z (-1458 y^3 + 2916 xyz - 972 z^2 + A)^{1/3}} \non \\
&& - {1 + i\sqrt{3} \over 6\cdot 2^{1/3} z} (-1458 y^3 + 2916 x y z - 972 z^2 
    +  A)^{1/3},
\end{eqnarray}
with $A=\sqrt{4(-81y^2+108xz)^3+(-1458y^3+2916xyz-972z^2)^2}$.
The cotangent bundle part (\ref{fm}) is led to the form:
\begin{eqnarray}
 \cH&=&-\oint_{-C^\prime}{d\xi \over 2\pi i}{\cF(1/\xi) \over \xi} {\partial \over \partial \xi}\ln\left(1+\xi x+{1 \over 4}\xi^2 y+{1 \over 36}\xi^3 z\right) \non \\
 &=&-\left({\cF(1/\xi_1) \over \xi_1}+{\cF(1/\xi_2) \over \xi_2}+{\cF(1/\xi_3) \over \xi_3}\right). \label{fr2}
 \end{eqnarray}
The result at an arbitrary point of $\Phi$ can be obtained by the following replacements
\begin{align}
x & \rightarrow~ (g^{-1})_i^{~j}\Psi_j\bar{\Psi}^i\,, \\
 {1 \over 4}y  & \rightarrow~ 
   {1 \over 2}((g^{-1})_i^{~j}\Psi_j\bar{\Psi}^i)^2
  -{1 \over 4}\tilde{R}_{i~k}^{~j~l}\bar{\Psi}^i\Psi_j\bar{\Psi}^k\Psi_l\,,  \\
 -{1 \over 36}z & \rightarrow -{1 \over 6}((g^{-1})_i^{~j}\Psi_j\bar{\Psi}^j)^3
+{1 \over 4}((g^{-1})_i^{~j}\Psi_j\bar{\Psi}^i)(\tilde{R}_{k~m}^{~l~~n}\bar{\Psi}^k\Psi_l\bar{\Psi}^m\Psi_n)\nonumber \\
 & ~~~~-{1 \over 12}|(g^{-1})_i^{~j}\tilde{R}_{j~l}^{~k~m}\Psi_k\bar{\Psi}^l\Psi_m|^2\,, \label{recover-co}
\end{align}
where $\tilde{R}_{i~k}^{~j~l}=(g^{-1})_i^{~m}(g^{-1})_n^{~j}(g^{-1})_k^{~p}(g^{-1})_q^{~l}R_{m~p}^{~~n~q}$. 

Note that for the $E_{7(-25)}/E_6\times U(1)$ case, we can construct the tangent bundle action by using (\ref{tbsol}), but it is technically difficult to perform the generalized Legendre transformation to obtain the cotangent bundle action. The situation is different from the $E_{6(-14)}/SO(10)\times U(1)$ case in Appendix \ref{GLT}. That is why the formula (\ref{fm}) for the cotangent bundle over any HSS is developed \cite{KuNo, AB}.

%%%%%%%%%%%%%%%%%%%%%%%%%%%%%%%%%%%%%%%%%%%%%%%%%
%
% Conclusion
%
%%%%%%%%%%%%%%%%%%%%%%%%%%%%%%%%%%%%%%%%%%%%%%%%%
\sect{Conclusion}
We have constructed the $\cN=2$ SUSY NLSMs on the cotangent bundles over the non-compact 
exceptional HSSs $\cM=E_{6(-14)}/SO(10)\times U(1)$ and $E_{7(-25)}/E_6 \times U(1)$
by using the results elaborated in \cite{KuNo} and \cite{AB}. 
The point is to use the projective superspace formalism which is an $\cN=2$ off-shell superfield formulation. 
Once an $\cN=1$ SUSY NLSM on a certain K\"ahler manifold is obtained, it is possible to extend it to the 
$\cN=2$ SUSY model containing the corresponding $\cN=1$ SUSY NLSM. 
We first have derived the transformation laws of the fields parameterizing $\cM$ and have constructed 
the $\cN=1$ SUSY NLSMs on $\cM$ invariant under the derived transformation laws. 
Second we have extended the $\cN=1$ SUSY NLSMs to ones with the $\cN=2$ SUSY models by using 
the explicit formula of the cotangent bundle over any HSS  developed in \cite{KuNo,AB}.
In this work, we complete constructing the K\"ahler potentials of the cotangent bundles 
over {\it all} the compact and non-compact HSSs listed in Table \ref{HSS}.

\vspace{5mm}

\noindent
{\bf Acknowledgements:}\\
The work of M.A. is supported in part by Grants-in-Aid for Scientific 
Research from the Ministry of Education, Culture, Sports, Science and Technology 
(No.25400280). \\

%%%%%%%%%%%%%%%%%%%%%%%%%%%%%%%%%%%%%%%%%%
%
% Appendix A
%
%%%%%%%%%%%%%%%%%%%%%%%%%%%%%%%%%%%%%%%%%%
\appendix
\noindent {\bf \Large Appendix}

\section{Clifford algebra} \label{appendix}
In this appendix,
we summarize the Clifford algebra for $SO(10)$ group \cite{DV, ItKuKu, KuSa}.
The Clifford algebra is generated by the $32\times 32$ gamma matrices
$\Gamma_{A}\,(A=1,\cdots,10)$ satisfying
\begin{equation}
\{\Gamma_{A},\Gamma_{B}\}=2\delta_{AB}{\bf 1}_{32}. \label{com}
\end{equation}
The matrices $\Gamma_{A}$ are hermitian:
\begin{equation}
\Gamma^{\dagger}_{A}=\Gamma_{A}.
\end{equation}
The complete set of the gamma matrices is spanned by $\{\Gamma^{(f)}\}(f=0,\cdots,10)$ given by
\begin{eqnarray}
&& \Gamma^{(f)}\equiv i^{[f/2]}\Gamma^{A_1A_2\cdots A_f}\equiv  i^{[f/2]}\Gamma^{[A_1}\Gamma^{A_2}\cdots\Gamma^{A_f]} \non \\
&&\quad\quad ={i^{[f/2]} \over n!}(\Gamma^{A_1}\Gamma^{A_2}\cdots\Gamma^{A_f}-\Gamma^{A_2}\Gamma^{A_1}\cdots\Gamma^{A_f}+\cdots), \\
&&(\Gamma^{(f)})^\dagger=\Gamma^{(f)},\quad (\Gamma^{(f)})^2={\bf 1}_{32},
\end{eqnarray}
where $[f/2]$ is the largest integer less than or equal to $f/2$ and $[A_1,\cdots,A_f]$ means
antisymmetrization.

By using the gamma matrices, we have
\begin{eqnarray}
 & \Gamma^{11}\equiv i\Gamma^1\Gamma^2\cdots \Gamma^{10}, & \\
 & (\Gamma^{11})^2={\bf 1}_{32}, &
\end{eqnarray}
by which we define the projection operator
\begin{eqnarray}
 P^{\pm}={1 \over 2}({\bf 1}_{32}\pm \Gamma_{11}). \label{app.eq.proj}
\end{eqnarray}

There exists a charge conjugation matrix which relates $\Gamma^A$ to $\Gamma^{A*}$. The latter forms an equivalent representation of the Clifford algebra:
\begin{eqnarray}
 \Gamma^A=-C^{-1}\Gamma^{A*}C. \label{gc}
\end{eqnarray}
The matrix $C$ has the following properties:
\begin{equation}\label{eq.charge}
C^{\mathrm{T}}=-C,\quad C^{\dagger}=C^{-1}.
\end{equation}
(\ref{gc}) is rewritten as
\begin{eqnarray}
 (\Gamma^A)^{\rm T}=-C\Gamma^AC^{-1}. \label{app.eq.gamma}
\end{eqnarray}
For $\Gamma^{11}$ and $P^{\pm}$, we have
\begin{eqnarray}
 &(\Gamma^{11})^{\rm T}=-C\Gamma^{11}C^{-1},& \\
 &(P^{\pm})^{\rm T}=CP^{\mp}C^{-1}.&
\end{eqnarray}

Let us give representation of the gamma matrix. We take the $\Gamma^{11}$ to be diagonal:
\begin{eqnarray}
 \Gamma^{11}=\begin{pmatrix}
   {\bf 1}_{16} & 0  \\
   0 & -{\bf 1}_{16} \\
 \end{pmatrix}.
\end{eqnarray}
In this case, the gamma matrices are block off-diagonal.
They are constructed by a tensor product of the
gamma matrices of $SO(6)$ and $SO(4)$ (for instance, see \cite{KuSa}):
\begin{eqnarray}
 \Gamma_A=\begin{pmatrix}
   0 & (\sigma_A)_{\a\b} \\
   (\sigma_A^\dagger)^{\a\b} & 0\\
 \end{pmatrix}, \quad \sigma_A^{\rm T}=\sigma_A,\quad A=1,\cdots, 10,\quad \a,\b=1,\cdots,16, \label{gamma}
\end{eqnarray}
where the $\sigma_A$'s are the gamma matrices on the Weyl spinor basis.
The generators of the SO(10) group are defined by
\begin{eqnarray}
 \sigma_{AB}={1 \over 4}(\sigma_A\sigma_B^\dagger-\sigma_A^\dagger\sigma_B),\quad
 \bar{\sigma}_{AB}={1 \over 4}(\sigma_A^\dagger\sigma_B-\sigma_A\sigma_B^\dagger). \label{sog}
\end{eqnarray}

The charge conjugation matrix $C$ takes the form
\begin{equation} \label{ccm}
C=\begin{pmatrix}
0            & \mathcal{C}\\
-\mathcal{C} & 0\\
\end{pmatrix},
\end{equation}
where $\mathcal{C}$ is the $16 \times 16$ matrix given by
\begin{equation}
\mathcal{C}=\begin{pmatrix}
0                                     & -\boldsymbol{1}_{4} \otimes i\sigma_{2}\\
\boldsymbol{1}_{4}\otimes i\sigma_{2} & 0
\end{pmatrix}=\mathcal{C}^{\rm T}=\mathcal{C}^{-1}=\mathcal{C}^\dagger.
\end{equation}

%%%%%%%%%%%%%%%%%%%%%%%%%%%%%%%%%%%%%%%%%%
%
% Appendix B
%
%%%%%%%%%%%%%%%%%%%%%%%%%%%%%%%%%%%%%%%%%%
\setcounter{equation}{0}
\section{Deriving (\ref{trans4}) from (\ref{trans3})}\label{calc}
To derive (\ref{trans4}) from (\ref{trans3}), we use the Fierz identity (for instance, see \cite{ke}). 
In the Dirac representation of $SO(10)$, the Fierz identity is described by
\begin{eqnarray}
\bar{\psi}^a\lambda_b={1 \over 32}\sum_{n=0}^5{a_n \over n!}(\Gamma_{A_1\cdots A_n})^a_{~b}(\lambda \Gamma^{A_1\cdots A_n}\bar{\psi}),\quad a,b=1,\cdots 32, &\label{i1}
\end{eqnarray}
where the indices $A_i(i=1,\cdots 5)$ run from 1 to 10 and 
$\bar{\psi}$, $\lambda$ are the Dirac spinors of $SO(10)$.
In this Appendix, we express the summation over the indices $A_i$ by contraction with upper and lower indices.
The coefficients $a_n$ are given by
\begin{eqnarray}
 a_0=a_1=a_4=2, \quad a_2=a_3=-2,\quad a_5=1.
\end{eqnarray}
For the gamma matrix there is a useful identity:
\begin{eqnarray}
\Gamma_{A_1\cdots A_n}\Gamma_{B_1\cdots B_m}\Gamma^{A_1\cdots A_n}=n!c(n,m)\Gamma_{B_1\cdots B_m}. \label{i2}
\end{eqnarray}
The value of the coefficient $c(n,m)$ is given in Table 2.
\begin{table}
\begin{center}
\begin{tabular}{c|cccccc}
\backslashbox{m}{n} & 0 & 1 & 2 & 3 & 4 & 5 \\ \hline
 0    & 1 & 10 & $-45$ & $-120$ & 210 & 252 \\
 1    & 1 & $-8$ & $-27$ & 48 & 42 & 0 \\
 2    & 1 & 6   & $-13$ & $-8$ & $-14$ & $-28$ \\
 3    & 1 & $-4$ & $-3$   & $-8$ & $-14$ & 0 \\
 4    & 1 & 2   & 3    &  8  & 2    & 12 \\
 5    & 1 & 0   & 5    &  0  & 10  & 0  
\end{tabular}
\caption{Coefficients $c(n,m)$} 
\end{center}
\end{table}
We also use the following identity:
\begin{eqnarray}
\psi_c\Gamma_{A_1A_2A_3}\psi=0, \label{i3a} 
\end{eqnarray}
where 
\begin{eqnarray}
 \psi_c=\psi^{\rm T}C. \label{c11}
\end{eqnarray}
It is easy to prove this identity by taking transposition of the left-hand side of (\ref{i3a}) and using (\ref{app.eq.gamma}).
The following identity is also useful:
\begin{eqnarray}
 \psi_c\Gamma_{A_1\cdots A_n}P^+\psi=0. \label{i3} 
\end{eqnarray}
This holds for any Dirac spinor of $SO(10)$ with $n=3$.
This is checked by transposition of the left-hand side of (\ref{i3}) with (\ref{app.eq.gamma}). 
On the other hand, for the Dirac spinor satisfying $\psi=P^+\psi$, (\ref{i3}) also holds for $n=2m$ with integer $m$. 
One can prove it by using (\ref{c11}) and (\ref{app.eq.gamma}).
The other useful identity for any Dirac spinor is
\begin{eqnarray}
 (\psi_c\Gamma_A^+\psi)(\bar{\lambda}^{\rm T}\Gamma^{+A}\psi)=0. \label{i4}
\end{eqnarray}
This is proved as follows. With the use of the Fierz identity (\ref{i1}), 
we have
\begin{eqnarray}
(\psi_c\Gamma_A^+\psi)(\bar{\lambda}^{\rm T}\Gamma^{+A}\psi)={1 \over 32}\sum_{n=0}^5  a_n(\psi_c\Gamma_{A_1\cdots A_n}\psi)(\bar{\lambda}^{\rm T}\Gamma_B^+\Gamma^{A_1\cdots A_n}\Gamma^{B+}\psi).
\end{eqnarray}
The first parenthesis in the right-hand side vanishes for $n=3$ because of (\ref{i3a}).
By using $\{\Gamma_A, \Gamma_{11} \}=0$ and (\ref{i2}), it is seen that
the last parenthesis in the right-hand side is not vanishing only for $n=1$ case.
Consequently in the summation $n=1$ part only remains to be nonzero and the right-hand side just gives the same form with the left-hand side multiplied by $-1/2$. It leads to (\ref{i4}).

Now let us apply these identities to (\ref{trans3}). 
Each term in the right-hand side in (\ref{trans3}) becomes
\begin{eqnarray}
 (\phi_c\Gamma_{AB}P^-\bar{\varepsilon}_c)(\Gamma^{AB}P^+\phi)_a
&=&{1 \over 32}{\Big \{}
-45a_0c(2,0)(\phi_cP^+\phi)(P^-\bar{\varepsilon}_c)_a \nonumber \\
&&-27a_1c(2,1)(\phi_c\Gamma_AP^+\phi)(\Gamma_AP^-\bar{\varepsilon}_c)_a \nonumber \\
&&+{a_5 \over 4!}c(2,5)(\phi_c\Gamma_{(5)}P^+\phi)(\Gamma_{(5)}P^-\bar{\varepsilon}_c)_a
{\Big \}}, \label{b1} \\
 (\phi_cP^-\bar{\varepsilon}_c)(P^+\phi)_a&=&{1 \over 32}
{\Big \{}
 a_0(\phi_cP^+\phi)(P^-\bar{\varepsilon}_c)_a+a_1(\phi_c\Gamma_AP^+\phi)(\Gamma_AP^-\bar{\varepsilon}_c)_a \nonumber \\
&&+{a_5 \over 5!}(\phi_c\Gamma_{(5)}P^+\phi)(\Gamma_{(5)}P^-\bar{\varepsilon}_c)_a
{\Big \}}, \label{b2}
\end{eqnarray}
where we have used (\ref{i2}).
Note that since $\phi=P^+\phi$, we have also used (\ref{i3}). 
Substituting (\ref{b1}) and (\ref{b2}) into (\ref{trans3}), it is seen that
the last terms in (\ref{b1}) and (\ref{b2}) cancel and (\ref{trans3}) finally
becomes (\ref{trans4}).

%%%%%%%%%%%%%%%%%%%%%%%%%%%%%%%%%%%%%%%%%%%%%%%%%
%
% Appendix C
%
%%%%%%%%%%%%%%%%%%%%%%%%%%%%%%%%%%%%%%%%%%%%%%%%%
\setcounter{equation}{0}
\section{Tangent bundle over $E_{6(-14)}/SO(10)\times U(1)$ and the Legendre transformation}\label{GLT}
In this section, we derive the tangent bundle action for the non-compact exceptional HSS $E_{6(-14)}/SO(10)\times U(1)$ by using the general formula (\ref{tbsol}).
First we need to calculate the first-order differential operator (\ref{fd}) which is for the $E_{6(-14)}/SO(10)\times U(1)$ case:
\begin{eqnarray}
 {\cal R}_{\Sigma,\bar{\Sigma}}=
-{1 \over 2}\Sigma_\a\bar{\S}^\b\Sigma_\g R^{\a~\g}_{~\b~\d}
 (g^{-1})^\d_{~\epsilon} 
 {\partial \over \partial \S_\epsilon}~. \label{fd-NCE}
\end{eqnarray}
Since a symmetric space is homogeneous, it is sufficient to 
perform the calculations of our interest at the origin, $\F=0$. 
By using the (\ref{m1}) and (\ref{r1}), the operator (\ref{fd-NCE}) is then written by
\begin{equation}
\cR_{\Sigma,\bar\Sigma}{\Big |}_{\F=\bar{\F}=0}= 
-|\S|^2\Sigma_{\beta}\frac{\partial}{\partial\Sigma_{\beta}}
+\frac{1}{4}\sum_A\bigl(\Sigma^{\rm T} \cC\sigma_A^{\dagger}\Sigma\bigr)(\sigma_A\cC^\dagger)_{\b\g}\bar\Sigma^{\g}\frac{\partial}{\partial\Sigma_{\b}}, \label{cR}
\end{equation}
where $|\S|^2=\Sigma_{\alpha}\bar\Sigma^{\alpha}$.

Let us calculate (\ref{tbsol}). 
Substituting (\ref{cR}) into (\ref{tbsol}), we have 
\begin{eqnarray}
{\cal L}(\Phi&=&\bar{\Phi}=0,\S,\bar{\S}) \non \\
&=&-g^\a_{~\b}\bar{\S}^\b{e^{\cR_{\S,\bar{\S}}}-1 \over \cR_{\S,\bar{\S}}}\S_\a{\Big |}_{\F=\bar{\F}=0} \non \\
&=&-\bar{\S}^\a\left(1+{1 \over 2}\cR_{\S,\bar{\S}}+{1 \over 3!}(\cR_{\S,\bar{\S}})^2+\cdots \right)\S_\a{\Big |}_{\F=\bar{\F}=0} \non \\
&=&-|\S|^2+{1 \over 2}|\S|^4
-{1 \over 8}|\S \cC\sigma_A^\dagger\S|^2 -{1 \over 3}|\S|^6
 +{1 \over 6}|\S|^2|\S \cC\sigma_A^\dagger\S|^2+\cdots, \label{c1}
\end{eqnarray}
where $|\S \cC\sigma_A^\dagger\S|^2=\sum_A(\bar{\S}\sigma_A\cC^\dagger \bar{\S})(\S \cC\sigma_A^\dagger\S)$. Here we have used the identity
\begin{eqnarray}
 \sum_A(\sigma_A^\dagger\S)(\S \cC\sigma_A^\dagger\S)=0. \label{i0}
\end{eqnarray}
This follows from the Fierz identity (\ref{i1}).
From the expression (\ref{c1}) we find the following form for the tangent space part
\begin{eqnarray}
{\cal L}(\Phi=\bar{\Phi}=0,\S,\bar{\S})
 =-\ln\left(1+|\S|^2+{1 \over 8}|\S^{\rm T}\cC\sigma_A^\dagger \S|\right).
\end{eqnarray}
One can extend this expression to one at an arbitrary point $\F$ of the base manifold by making the replacement
\begin{eqnarray}
 |\S|^2\rightarrow g^\a_{~\b}\S_\a\bar{\S}^\b, \quad
 {1 \over 8}|\S^{\rm T}\cC\sigma_A^\dagger \S|^2\rightarrow {1 \over 2}(g^\a_{~\b}\S_\a\bar{\S}^\b)^2-{1 \over 4}R^{\a~\g}_{~\b~\d}\S_\a\bar{\S}^\b\S_\g\bar{\S}^\d.
\end{eqnarray}
Then we obtain the action of the tangent space part
\begin{eqnarray}
{\cal L}(\Phi,\bar{\Phi},\S,\bar{\S})
=-\ln\left(
1+g^\a_{~\b}\S_\a\bar{\S}^\b+{1 \over 2}(g^\a_{~\b}\S_\a\bar{\S}^\b)^2
-{1 \over 4}R^{\a~\g}_{~\b~\d}\S_\a\bar{\S}^\b\S_\g\bar{\S}^\d
\right). \label{tb}
\end{eqnarray}
This is the correct result for $\cL$. 
Indeed, one can check that (\ref{tb}) satisfies the equation (\ref{fd2}) 
which reads in the present case,
\begin{eqnarray}
{1 \over 2}R^{\a~\g}_{~\b~\d}(g^{-1})^\delta_{~\epsilon}
{\partial \cL \over \partial \S_\epsilon}\S_\a\S_\g
 +{\partial \cL \over \partial \bar{\S}^\b}+g^\a_{~\b}\S_\a=0. \label{m}
\end{eqnarray}
Let us briefly prove that (\ref{tb}) satisfies this equation. It is again sufficient to consider at $\F=0$. In this case, the first term of the left-hand side (\ref{m}) becomes
\begin{eqnarray}
&&{1 \over 2}R^{\a~\g}_{~\b~\d}(g^{-1})^\delta_{~\epsilon}{\partial \cL \over \partial \S_\epsilon}\S_\a\S_\g{\Big |}_{\F=\bar{\F}=0}
\non \\
&&\quad =-{1 \over Z}\left(
2\S_\b|\S|^2+{1 \over 4}\S_\b|\S^{\rm T} \cC\sigma_A^\dagger \S|^2-{1\over 2}\sum_A(\sigma_A \cC^\dagger\bar{\S})_\b(\S^{\rm T} \cC\sigma_A^\dagger \S)\right),  \\
&&Z\equiv1+|\S|^2+{1 \over 8}|\S^{\rm T} \cC\sigma_A^\dagger \S|^2,
\end{eqnarray}
where we have used (\ref{i0}). This exactly cancels against other terms in (\ref{m}).

Next we derive the $\cN=2$ SUSY NLSM on the cotangent bundle over 
$E_{6(-14)}/SO(10)\times U(1)$ from (\ref{tb}) by using the Legendre transformation.
We consider the first order action (\ref{f-o}), which in the present case is written by
\begin{eqnarray}
 S=\int {\rm d}^8z{\Big \{}K(\F,\bar{\F})
 &-& \ln\left(1+g^\a_{~\b}U_\a\bar{U}^\b+{1 \over 2}(g^\a_{~\b}U_\a\bar{U}^\b)^2
 -{1 \over 4}R^{\a~\g}_{~\b~\d}U_\a\bar{U}^\b U_\g\bar{U}^\d\right)
 \non \\
 &+&U_\a \J^\a + \bar{U}^\a \bar{\J}_\a{\Big \}}.
 \label{parent}
\end{eqnarray}
Let us eliminate $U$ and $\bar{U}$.
As in the case of the tangent bundle, we again consider the action at $\F=0$ 
since the base manifold is homogeneous. Then, the action (\ref{parent}) is
\begin{eqnarray}
 &\displaystyle
  S=\int d^8z \left(-\ln\Omega+U_\a \psi^\a +
  \bar{U}^{\a}\bar{\psi}_{\a}\right), 
 \label{cal3} &
\end{eqnarray}
where $\psi$ is a cotangent vector at $\Phi=0$ and
\begin{eqnarray}
  \O=1+|U|^2+{1 \over 8}|U^{\rm T} \cC \s_A^\dagger  U|^2, \label{defo}
\end{eqnarray}
with
\begin{eqnarray}
&|U|^2=\bar{U}^{\rm T}U, & \\
& |U^{\rm T} \cC\sigma_A^\dagger U|^2=\sum_A(U^{\rm T}\cC\sigma_A^\dagger U)(\bar{U}^{\rm T}\sigma_A \cC^\dagger \bar{U}).&
\end{eqnarray}
The equations of motion for $U$ and $\bar{U}$ are
\begin{eqnarray}
& \displaystyle{U_\a+\sum_A(\s_A \cC^\dagger \bar{U})_\a (U^{\rm T}\cC\s_A^\dagger U)/4 \over
  \O}=\bar{\psi}_\a, \label{pp1} & \\
&  \displaystyle{\bar{U}^\a+\sum_A(\cC\s_A^\dagger U)^\a (\bar{U}^{\rm T}\s_A \cC^\dagger \bar{U})/4 \over \O}=\psi^\a. \label{pp2} &
\end{eqnarray}
These equations yield
\begin{eqnarray}
 & \displaystyle \bar{\psi}^{\rm T}\cC\s_A^\dagger \bar{\psi}={U^{\rm T}\cC\s_A^\dagger U \over \O},& \\
 & \displaystyle \psi^{\rm T}\s_A \cC^\dagger \psi = {\bar{U}^{\rm T}\s_A \cC^\dagger \bar{U} \over \O},& \\
&\bar{\psi}^{\rm T}\psi\equiv \displaystyle |\psi|^2={1 \over \Omega^2}\left(
 |U|^2+{1 \over 2}|U^{\rm T}\cC\sigma_A^\dagger U|^2
  +{1 \over 8}|U|^2|U^{\rm T}\cC\sigma_A^\dagger U|^2
\right). &
\end{eqnarray}
From these equations we have
\begin{eqnarray}
& \displaystyle
{1 \over 4}-|\psi|^2+{1 \over 2}|\psi^{\rm T} \s_A \cC^\dagger \psi|^2
=\left({1 \over 2}-{|U|^2 \over \O}\right)^2,
 \label{sol-a} &
\end{eqnarray}
where $|\psi^{\rm T}\cC\sigma_A^\dagger \psi|^2=\sum_A(\psi^{\rm T}\cC\sigma_A^\dagger \psi)(\bar{\psi}^{\rm T}\sigma_A \cC^\dagger \bar{\psi})$.
From (\ref{defo}) it is seen that the correspondence between the tangent and cotangent
vectors should be such that $U\rightarrow 0 \Leftrightarrow
\psi\rightarrow 0$. 
This means that we have to choose the following solution of (\ref{sol-a}).
\begin{eqnarray}
 {|U|^2 \over \O}={1 \over 2}-\sqrt{{1\over 4}-|\psi|^2
+{1\over 2}|\psi^{\rm T} \s_A \cC^\dagger \psi|^2}~.
\end{eqnarray}
From the above result we obtain $\O$ in terms of $\psi$ and its conjugate.
By definition of $\Omega$ (\ref{defo}), we have
\begin{eqnarray}
 {1 \over \O}={1 \over \O^2}+{|U|^2 \over \O^2}
+{1\over 8}\left|{\psi^{\rm T} \s_A \cC^\dagger \psi \over \O}\right|^2.
\end{eqnarray}
This is equivalent to 
\begin{eqnarray}
\left({1 \over \O}-{\Lambda \over 2}\right)^2={\Lambda^2 \over 4}
 -{1 \over 8}|\psi^{\rm T} \s_A \cC^\dagger \psi|^2, 
  \label{sol-b}
\end{eqnarray}
where
\begin{eqnarray}
 \Lambda={1 \over 2}+\sqrt{{1\over 4}-|\psi|^2
 +{1\over 2}|\psi^{\rm T} \s_A \cC^\dagger \psi|^2}.
\end{eqnarray}
Since we should have $\O\rightarrow 1$ when $\psi\rightarrow 0$, it is
necessary to choose the following solution for (\ref{sol-b}):
\begin{eqnarray}
 {1 \over \O}={\Lambda \over 2}+\sqrt{{\Lambda^2\over 4}
 -{1\over 8}|\psi^{\rm T} \s_A \cC^\dagger \psi|^2}
 ={\Lambda \over 2}+{1 \over 2}\sqrt{\Lambda-|\psi|^2}. \label{O}
\end{eqnarray}
The above result is of one at the origin $\F=0$ of the base manifold.
In order to extend to one at an arbitrary point $\F\neq 0$, we have to make
the following replacement:
\begin{eqnarray}
 &|\psi|^2 ~\rightarrow~
  (g^{-1})^\a_{~\b}\J^\b\bar{\J}_\a,&\nonumber \\
 & \displaystyle {1\over 8}|\psi^{\rm T} \s_A \cC^\dagger \psi|^2 ~\rightarrow ~
{1 \over
  2}((g^{-1})^\a_{~\b}\J^\b\bar{\J}_\a)^2
 -{1 \over 4}\tilde{R}^{\a~\g}_{~\b~\d}\bar{\J}_\a\J^\b\bar{\J}_\g\J^\d,&
\end{eqnarray}
where
$\tilde{R}^{\a~\g}_{~\b~\d}=(g^{-1})^{\a}_{~\a^\prime}(g^{-1})^{\b^\prime}_{~\b}
 (g^{-1})^{\g}_{~\g^\prime}(g^{-1})^{\d^\prime}_{~\d}
 R^{\a^\prime~\g^\prime}_{~\b^\prime~\d^\prime}.$
Substituting (\ref{pp1}), (\ref{pp2}) and (\ref{O}) into (\ref{parent}), we obtain the cotangent bundle action $S_{\rm ctb}$
\begin{eqnarray}
 S_{\rm ctb}=\int d^8z (K(\F,\bar{\F})+{\cal H}(\F,\bar{\F},\J,\bar{\J})),
\end{eqnarray}
where
\begin{eqnarray}
 {\cal H}(\F,\bar{\F},\J,\bar{\J})&=&
 \ln\left(\L+\sqrt{\L-(g^{-1})^\a_{~\b}\J^\b\bar{\J}_\a}\right)\nonumber \\
 &&-2\L+{4((g^{-1})^\a_{~\b}\J^\b\bar{\J}_\a)^2
 -2\tilde{R}^{\a~\g}_{~\b~\d}\bar{\J}_\a\J^\b\bar{\J}_\g\J^\d
 \over \L+\sqrt{\L-(g^{-1})^\a_{~\b}\J^\b\bar{\J}_\a}}, \label{cot-E6}
\end{eqnarray}
and
\begin{eqnarray}
 \L={1 \over 2}+\sqrt{{1 \over 4}-(g^{-1})^\a_{~\b}\J^\b\bar{\J}_\a
    +2((g^{-1})^\a_{~\b}\J^\b\bar{\J}_\a)^2
    -\tilde{R}^{\a~\g}_{~\b~\d}\bar{\J}_\a\J^\b\bar{\J}_\g\J^\d}.
\end{eqnarray}
The cotangent space part (\ref{cot-E6}) coinsides with  (\ref{fr1}) up to the irrelevant constant, which is a conseuquence of deformation of the contour from $C$ to $-C^\prime$ in (\ref{fm}).

Finally let us check that the cotangent bundle action (\ref{cot-E6}) satisfies the
equation (\ref{ce}). In the present case, the equation takes the following form:
\begin{eqnarray}
 \S_\a g^\a_{~\b}-{1 \over 2}\S_\a\S_\g
  R^{\a~\g}_{~\b~\d}(g^{-1})^\d_{~\e}\J^\e
 =\bar{\J}_\b. 
 \label{cot-fd}
\end{eqnarray}
To prove this, we again set $\F=0$.
Then, the left-hand side in (\ref{cot-fd}) becomes
\begin{eqnarray}
&&\S_\b-{1 \over 2}\S_\a\S_\g
R^{\a~\g}_{~\b~\d}(g^{-1})^\d_{~\e}\J^\e{\bigg |}_{\F=0} \\
&&\quad \quad =\S_\b-{1 \over 2}\left(
  2(\S_\a \psi^\a)\S_\b-{1 \over 2}\sum_A(\s_A \cC^\dagger \psi)_\b(\S^{\rm
  T}\cC \s_A^\dagger \S)
\right) \non \\
&&\quad \quad ={1 \over \O}\left(\S_\b+{1 \over 4}\sum_A(\s_A \cC^\dagger \bar{\S})_\b
	       (\S^{\rm T}\cC \s_A^\dagger \S)\right), \non
\label{cal4}
\end{eqnarray}
where in the second equality we have used (\ref{pp2}) to express $\psi$ in terms of $\S$.
Taking (\ref{defo}) into account, we see that the expression obtained is exactly $\bar{\J}_\b$ at $\F=0$.

%%%%%%%%%%%%%%%%%%%%%%%%%%%%%%%%%%%%%%%%%%%%%%%%%
%
% Appendix D
%
%%%%%%%%%%%%%%%%%%%%%%%%%%%%%%%%%%%%%%%%%%%%%%%%%
\setcounter{equation}{0}
\section{Tangent bundle over $E_{7(-25)}/E_6\times U(1)$}
In this Appendix, we derive the tangent bundle action with the use of the formula (\ref{tbsol}).
First we calculate the differential operator (\ref{fd}), which is written in the present case
\begin{eqnarray}
 \cR_{\S,\bar{\S}}=-{1 \over 2}\S^i\bar{\S}_j\S^kR_{i~k}^{~j~l}(g^{-1})_l^{~m}{\partial \over \partial \S^m}, \label{do7}
\end{eqnarray}
where $(g^{-1})_{~i}^{j}=(g^{~j}_{i})^{-1}$ is the inverse metric of $g^{~j}_{i}$. 
We set $\phi=\bar{\phi}=0$ as in the $E_{6(-14)}/SO(10)\times U(1)$ case.
Substituting (\ref{m2}) and (\ref{r2}) into (\ref{do7}), we have
\begin{eqnarray}
 \cR_{\S,\bar{\S}}{\Big |}_{\phi=\bar{\phi}=0}=-|\S|^2\S^j{\partial \over \partial \S^j}+{1 \over 2}(\Gamma_{mik}\S^i\S^k)\Gamma^{mjl}\bar{\S}_j{\partial \over \partial \S^l},\quad |\S|^2:=\S^i\bar{\S}_i.
\end{eqnarray}

We are ready to calculate the tangent space part $\cL$. Starting with (\ref{tbsol}), we have
\begin{eqnarray}
&&\cL(\phi=\bar{\phi}=0,\S,\bar{\S})=-{e^{\cR_{\S,\bar{\S}}}-1 \over \cR_{\S,\bar{\S}}}g_i^{~j}\bar{\S}_j\S^i{\Big |}_{\phi=\bar{\phi}=0} \non \\
&& \quad =-\left(1+{1 \over 2}\cR_{\S,\bar{\S}}+{1 \over 3!}(\cR_{\S,\bar{\S}})^2+\cdots \right)g_i^{~j}\bar{\S}_i\S^j{\Big |}_{\phi=\bar{\phi}=0} \non \\
&& \quad = -|\S|^2+\hf |\S|^4-{1 \over 4}(\Gamma_{ijk}\Sigma^j\S^k)(\Gamma^{ilm}\bar{\S}_l\bar{\S}_m)-{1 \over 3}|\Sigma|^6 \non \\
&& \quad \quad +{1 \over 4}|\S|^2(\Gamma_{ijk}\Sigma^j\S^k)(\Gamma^{ilm}\bar{\S}_l\bar{\S}_m) \non \\
&& \quad \quad +{1 \over 36}(\Gamma_{ijk}\S^i\S^j\S^k)(\Gamma^{lmn}\bar{\S}_l\bar{\S}_m\bar{\S}_n)+\cdots. 
\label{tb-ex}
\end{eqnarray}
From (\ref{tb-ex}), we conjecture the form of the tangent space part
\begin{eqnarray}
 &&\cL(\phi=\bar{\phi}=0,\S,\bar{\S}) \\
 &&~~=-\ln\left(1+|\S|^2+{1 \over 4}(\Gamma_{ijk}\Sigma^j\S^k)(\Gamma^{ilm}\bar{\S}_l\bar{\S}_m)+{1 \over 36}(\Gamma_{ijk}\S^i\S^j\S^k)(\Gamma^{lmn}\bar{\S}_l\bar{\S}_m\bar{\S}_n)\right). \non
\end{eqnarray}
The tangent space part at an arbitrary point of the base manifold is obtained by the following replacements:
\begin{eqnarray}
 |\S|^2 & \rightarrow & g_i^{~j}\S^i\bar{\S}_j, \\
{1 \over 4}(\Gamma_{ijk}\Sigma^j\S^k)(\Gamma^{ilm}\bar{\S}_l\bar{\S}_m) & \rightarrow & {1 \over 2}(g_i^{~j}\S^i\bar{\S}_j)^2-{1 \over 4}R_{i~k}^{~j~l}\S^i\bar{\S}_j\S^k\bar{\S}_l, \\
{1 \over 36}(\Gamma_{ijk}\S^i\S^j\S^k)(\Gamma^{lmn}\bar{\S}_l\bar{\S}_m\bar{\S}_n) & \rightarrow & {1 \over 6}(g_i^{~j}\S^i\bar{\S}_j)^3-{1 \over 4}(g_i^{~j}\S^i\bar{\S}_j)(R_{k~l}^{~m~n}\S^k\bar{\S}_m\S^l\bar{\S}_n) \non \\
&& +{1 \over 12}|g_i^{~j}R_{j~l}^{~k~m}\bar{\S}_k\S^l\bar{\S}_m|^2,
\end{eqnarray}
which lead to
\begin{eqnarray}
 \cL(\phi,\bar{\phi},\S,\bar{\S})&=&-\ln\left(1+g_i^{~j}\S^i\bar{\S}_j+{1 \over 2}(g_i^{~j}\S^i\bar{\S}_j)^2
  -{1 \over 4}R_{i~k}^{~j~l}\S^i\bar{\S}_j\S^k\bar{\S}_l\right. \non \\
&&+{1\over 6}(g_i^{~j}\S^i\bar{\S}_j)^3-{1 \over 4}(g_i^{~j}\S^i\bar{\S}_j)(R_{k~l}^{~m~n}\S^k\bar{\S}_m\S^l\bar{\S}_n)  \non \\
&&+\left.{1 \over 12}|g_i^{~j}R_{j~l}^{~k~m}\bar{\S}_k\S^l\bar{\S}_m|^2 \right).
\end{eqnarray}

For the consistency check, we can prove that this satisfies the equation (\ref{fd2}), which is in the present case
\begin{eqnarray}
 {1 \over 2}R_{i~k}^{~~j~l}(g^{-1})_l^{~m}{\partial \cL \over \partial \S^m}\S^i\S^k+{\partial \cL \over \partial \bar{\S}_j}+g_{i}^{~j}\S^i=0. \label{con-e1}
\end{eqnarray}
The first term in (\ref{con-e1}) at $\phi=\bar{\phi}=0$ is calculated as
\begin{eqnarray}
&&{1 \over 2}R_{i~k}^{~~j~l}(g^{-1})_l^{~m}{\partial \cL \over \partial \S^m}\S^i\S^k{\Big |}_{\phi=\bar{\phi}=0}\non \\
&&\quad \quad=-{1 \over 2Z}{\Big (}2|\S|^2\S^j-(\Gamma_{mik}\S^i\S^k)\Gamma^{mjl}\bar{\S}_l \non \\
&&\quad \quad \quad  +{1 \over 18}\S^j|\Gamma_{ilm}\S^l\S^m|^2-{1 \over 6}(\Gamma^{jpq}\bar{\S}_p\bar{\S}_q)(\Gamma_{klm}\S^k\S^l\S^m)
 {\Big )}, \\
 && Z\equiv 1+|\S|^2+{1 \over 4}|\Gamma_{ijk} \S^j\S^k|^2+{1 \over 36}|\Gamma_{ijk}\S^i\S^j\S^k|^2,
\end{eqnarray}
where we have used the Springer relation (\ref{sp}). 
One can see that this term exactly cancels against the rest of terms in (\ref{con-e1}).
It should be emphasized that it is technically difficult to perform the Legendre transformation to obtain 
the cotangent bundle action as in the $E_{6(-14)}/SO(10)\times U(1)$. Therefore, we have used the formula (\ref{fm}) to
obtain the cotangent bundle action instead of using the Legendre transformation.

%%%%%%%%%%%%%%%%%%%%%%%%%%%%%%%%%%%%%%%%%%
%
% bibliography
%
%%%%%%%%%%%%%%%%%%%%%%%%%%%%%%%%%%%%%%%%%%

\end{document}